\pdfoutput=1
\documentclass[11pt,aps,a4paper,eqsecnum,amsmath,amssymb,superscriptaddress,notitlepage,nofootinbib,longbibliography]{revtex4-1}

\makeatletter
\def\@bibdataout@aps{%
\immediate\write\@bibdataout{%
@CONTROL{%
apsrev41Control%
\longbibliography@sw{%
    ,author="08",editor="1",pages="1",title="0",year="1"%
    }{%
    ,author="08",editor="1",pages="1",title="",year="1"%
    }%
  }%
}%
\if@filesw \immediate \write \@auxout {\string \citation {apsrev41Control}}\fi 
}
\makeatother

\usepackage{graphicx}
\usepackage{hyperref}
\usepackage{color}
\DeclareMathAlphabet{\mathpzc}{OT1}{pzc}{m}{it}

\newcommand{\repp}{\mathtt{p}}
\newcommand{\repq}{\mathtt{q}}
\newcommand{\defq}{{q}}
\newcommand{\ospq}{U_\defq[OSp(3|2)]}
\newcommand{\xxzq}{U_\defq[SU(2)]}

\newcommand{\eq}{\begin{equation}}
\newcommand{\en}{\end{equation}}

\newcommand{\bear}{\begin{eqnarray}}
\newcommand{\ear}{\end{eqnarray}}

\begin{document}
\preprint{}
\title{$\ospq$ quantum chains with quantum group invariant boundaries}  

\author{Holger Frahm}
\affiliation{%
Institut f\"ur Theoretische Physik, Leibniz Universit\"at Hannover,
Appelstra\ss{}e 2, 30167 Hannover, Germany}

\author{M\'arcio J. Martins}
\affiliation{%
Departamento de F\'isica, Universidade Federal de S\~ao Carlos,
C.P. 676, 13565-905 S\~ao Carlos (SP), Brazil}

\date{\today}

\begin{abstract}
Based on the finite-size analysis of the spectrum of the quantum group invariant deformation of the $OSp(3|2)$ superspin chain we identify the operator content of the conformal field theory describing the model in its scaling limit.  We find that the macroscopic degeneracy of the conformal weights observed in the thermodynamic limit  of the  isotropic superspin chain is lifted by the deformation.
\end{abstract}


\maketitle
\section{Introduction}
During the recent past years there has been an increasing activity on
the study of families of massless one-dimensional integrable models
with intricate critical behaviour similar to that expected for
conformal field theories (CFTs) based on non-compact symmetries. A
common property among of these lattice models is the presence of
towers of lower energy states leading to the same scaling dimension in
the thermodynamic limit whose degeneracies is typically lifted by
logarithmic corrections on the finite interval.  It has been argued
that examples of such systems are not restricted to specific models
but instead range from staggered two-dimensional vertex models to
families of spin chains based on twisted Lie algebras and as well as
on supergroup symmetries
\cite{EsFS05,IkJS08,FrMa11,FrMa12,VeJS14,VeJS16a,FrHM19}.
Probably best understood among these models is the staggered
six-vertex model \cite{IkJS08,IkJS12,CaIk13,FrSe14,BKKL19,BaKL21}:
here the spectrum of conformal weights has been shown to have
continuous components arising from the non-compact target space of the
underlying CFT.  Based on finite size studies probing properties of
the algebra of extended conformal symmetry the low energy effective
theory of this model has been argued to be a Lorentzian black hole
nonlinear sigma model \cite{BaKL21,BKKL21a}.

For conventional conformal field theories it is well known that
boundary conditions are able to select some of the conformal
dimensions predicted for the operator content of the underlying
conformal field theory from symmetry considerations.  Therefore, a
natural question to ask is whether or not the presence of towers of
low energy states and the presence of continuous components in the
finite-size spectrum is robust against changing the system boundary
conditions. Indeed, this kind of question has recently been
investigated in the context of the staggered six-vertex model
\cite{RoJS19,RoJS21,FrGe22,FrGe23}: in this case fine-tuning of the
boundary conditions is necessary for the low energy degrees of freedom
to be described by a non-compact boundary CFT.  Moreover, based on
numerical work a boundary RG flow to a fixed point with discrete
conformal spectrum has been identified.

The purpose of this work is to begin an investigation about the effect
of boundary condition on the operator content of the $\defq$-deformed
$OSp(3|2)$ superspin chain. In the isotropic $\defq \rightarrow 1$
limit this model is a lattice realization of the Goldstone phase of
intersecting loops and it has been pointed out that, for both periodic
and free boundary conditions, there exist several towers of states
with integer conformal dimensions subject to strong subleading
logarithmic corrections to scaling in the finite system
\cite{MaNR98,JaRS03,FrMa15,FrMa22}. On the other hand, it has been
observed that some of the towers are eliminated in the integrable
anisotropic deformation of the periodic model \cite{FrHM19}. At the
same time the corresponding corrections to scaling vanish as power
laws depending on the deformation parameter $\defq$ rather than
logarithmic in the thermodynamic limit.  This behaviour has been
linked to the observation that part of the energies of the
$\defq$-deformed $OSp(3|2)$ superspin chain coincides with the
eigenvalues of the integrable $XXZ$ spin $S=1$ chain for a special
value of the deformation parameter $\defq$.

Here we will study the critical properties of such $\defq$-deformed
superspin chain with boundary conditions restoring the full $\ospq$
quantum group invariance.  This is the simplest boundary condition
retaining part of the spectral degeneracies of the isotropic model and
at the same time recovers the free boundary conditions in the
isotropic limit $\defq\to1$.  In addition, the $\ospq$ spin chain is
solvable by the Bethe ansatz which allows allows to study its spectral
properties for very large systems.  For our analysis of the critical
properties we exploit the relationship between the eigenenergies
$E_n(L)$ of the finite system and the central charge $c$ and the
conformal weights $h_n$ (appearing as surface exponents describing the
asymptotic behaviour of boundary correlation functions) of the
corresponding operators in the boundary CFT describing the model in
the scaling limit \cite{BlCN86,ABBBQ87},
\begin{equation}
\label{eq:finitesize}
E_n(L) = L \varepsilon_{\infty} +f_{\infty} - \frac{\pi v_F}{24L}\,c +  \frac{\pi v_F}{L}\, h_n
	+ o\left(\frac{1}{L}\right)\,,
\end{equation}
where $\varepsilon_{\infty}$ is the energy per site of the ground
state, $f_{\infty}$ is the surface energy resulting from the open
boundary conditions, and $v_F$ is the Fermi velocity of the massless
low-lying excitations.
A peculiar feature of $\ospq$-invariant model is that the ground state
energy has no finite-size corrections beyond the surface energy
$f_\infty$ implying a vanishing central charge $c$ for generic values
of the deformation parameter $\defq$ in the massless regime.  This has
to be contrasted with other integrable quantum group invariant spin
chains such as those based on the spin-$s$ representation of the
$\xxzq$ quantum algebra in which the ground state finite-size
corrections only vanish for a specific choice of $\defq$
\cite{ABBBQ87,Mart90}.

Given the quantum group symmetry the eigenenergies of the superspin
chain appear in the multiplets of $OSp(3|2)$.  Following Ref.\
\cite{Jeugt84} they may be labelled by two indices $(\repp ;\repq)$.
Except for the trivial representation $(0;0)$ the index $\repp$ takes
nonnegative integer values while $\repq\geq \frac12$ is an integer or
half integer -- as in the isotropic case.  The presence of these
levels in the spectrum of the superspin chain of length $L$ requires
the selection rule $L-\repp-2\repq \in 2\mathbb{N}$ to be
satisfied. An exception are the 'atypical' representations such as
$(1;1)$ which appears in the Hilbert space of chains of any length as
part of reducible but indecomposable representations.

Combining exact diagonalization for small systems with the numerical
solution of the Bethe equations we have investigated a significant
number of low energy states in the sectors with different values of
the two $U(1)$ charges of the $\ospq$ superspin chain.
Using (\ref{eq:finitesize}) with data for sufficiently large systems
we provide strong evidence that the conformal weights in terms of the
quantum numbers $(\repp;\repq)$ are given as
\begin{equation}
\label{OPE}
\begin{aligned}
    h_{(\repp;\repq)} &= \frac{\pi-\gamma}{2\pi}\, \repp(\repp+1) + \frac\gamma{2\pi}\,2\repq(2\repq-1)\,
\end{aligned}
\end{equation}
in the massless region $0 \leq \gamma \leq \frac{\pi}{2}$. The
anisotropic parameter $\gamma$ is related to the quantum group
deformation by $\defq=e^{i\gamma/2}$.  By way of contrast with the
periodic case studied in \cite{FrHM19} we find no trace of the
presence of towers of states in the eigenspectrum for the quantum
group boundary condition. We anticipate however that some of the low
lying states have strong power law corrections to finite-size scaling
which required the study of large system sizes. In any case this
result further elucidates our earlier findings for the isotropic model
\cite{FrMa15,FrMa22}: (i) as $\gamma \rightarrow 0$ the second term of
Eq.(\ref{OPE}) vanishes and we are left with the conformal weights
identified in the finite-size analysis of the periodic chain. (ii) For
the isotropic model with free boundary conditions (\ref{OPE}) results
in a tower of states with integer conformal dimensions labelled on the
finite lattice by the index $\repp$.  Since the number of available
values for $\repq$ grows with the lattice size this leads to a
macroscopic degeneracy of the low energy states which is lifted by
corrections to scaling vanishing as $1/\log L$ in the thermodynamic
limit.

This paper is organized as follows: in the following section we
formulate the integrable $\ospq$-invariant superspin chain based on
particular solutions of the boundary Yang-Baxter algebra
\cite{Cher84,Skly88}.  Then we present the solution to this spectral
problem of the model in terms of Bethe equations.  Being based on a
superalgebra there exist two such solutions corresponding to different
orderings of the fermionic and bosonic states in the local basis.  In
Section~\ref{sec:xxzq} we uncover an exact correspondence between the
spectra of the $\defq$-deformed $OSp(3|2)$ superspin chain and the
$XXZ$ spin-$1$ spin chain, both with quantum-group invariant boundary
conditions, for a particular value of the deformation parameter
$\defq$.  Together with our finite-size analysis of the
$\ospq$-symmetric super spin chain this correspondence supports our
proposal (\ref{OPE}) for the operator content of the boundary CFT
describing the critical point of the model.  We close with a
discussion putting the present results in the context of earlier work
on the $OSp(3|2)$ superspin chain and its deformations as well as on
the influence of boundary conditions in other integrable spin chains
with a continuous component to the conformal spectrum.  The
$\ospq$-invariant Hamiltonian in terms of the $OSp(3|2)$ superalgebra
is presented in Appendix~\ref{app:HAMq}.

\section{Formulation of the model}

In this section we describe the integrable $\defq$-deformed $OSp(3|2)$
spin chain with quantum group invariance. We recall that the main
tools for dealing with open boundary conditions within the the quantum
inverse scattering method have been introduced by Cherednik and
Sklyanin \cite{Cher84,Skly88}. This pioneering work was further
elaborated and in particular it has been shown how integrable quantum
group invariant models can be constructed for a variety of distinct
affine Lie algebras in \cite{MeNe91a,KuSk91}.  Later on similar
constructions have been pursued to include integrable models with
underlying graded affine superalgebras \cite{LiGo96,BGZZ98}.
Specifically, the right and left boundary conditions are encoded in
reflection matrices $K^\pm(\lambda)$ satisfying the so-called boundary
Yang-Baxter equations, e.g.\
\begin{eqnarray}
	R_{12}(\lambda-\mu ) K_{1}^{-}(\lambda ) R_{21}(\lambda+\mu)
	K_{2}^{-}(\mu ) =  
	K_{2}^{-}(\mu )
	R_{12}(\lambda +\mu ) K_{1}^{-}(\lambda ) R_{21}(\lambda -\mu)\,
\label{REYB}
\end{eqnarray}
for the left boundary matrix $K^-(\lambda)$. Here $K^-_{j}(\lambda)$
is a copy of the reflection matrix acting non-trivially on the space
$V_j$ and $R_{ij}(\lambda) \in \text{End}\left(V_i\otimes V_j\right)$
is the $R$-matrix of the $\defq$-deformed $OSp(3|2)$ vertex model in
the fundamental representation for which $V_j$ is a five-dimensional
$\mathbb{Z}_2$-graded vector space \cite{GaMa04}.
The $R$-matrix satisfies the commutation property
$[\check{R}_{12}(\lambda), \check{R}_{12}(\mu)]=0$ where
$\check{R}_{12}(\lambda) = P_{12}\,R_{12}(\lambda)$ with the graded
permutation operator
$P_{12} \, a\otimes b= (-1)^{p_{a}p_{b}} b\otimes a$ on $V\otimes V$.%
\footnote{This is because $\check{R}$ belongs to the algebra generated
  by a braid operator and can therefore be obtained by means of the
  Baxterization method \cite{Jones90}.}  Recall here, that $p_a$
stands for the grading (or parity), i.e.\ $p_a=0$ if $a$ is an even
(bosonic) state and $p_a=1$ if $a$ is odd (fermionic).  As a
consequence it is easy to see that (\ref{REYB}) is satisfied by the
trivial solution $K^-(\lambda)\equiv \mathbf{1}$.  Given this solution
of (\ref{REYB}) $K^+(\lambda)$ is easily obtained using the crossing
properties of the $R$-matrix \cite{Skly88}.
Since the explicit form of the $R$-matrix depends on the choice of the
$\mathbb{Z}_2$-grading the same is true for the corresponding
$K^+(\lambda)$.  For instance, ordering the elements of the five
dimensional basis in the $fbbbf$ grading we obtain,
\begin{equation}
K^{+}(\lambda)=\left( \begin{array}{ccccc} 
	\defq^2 &  0 & 0 & 0 & 0   \\
	0 &  \defq^2 & 0 & 0 & 0  \\
	0 &  0 & \defq & 0 & 0  \\
	0 &  0 & 0 & 1 & 0  \\
	0 &  0 & 0 & 0 & 1  \\
	\end{array}
	\right)\,.
\end{equation}

With these objects the double row transfer matrix generating commuting
integrals of motion acting on the Hilbert space $\otimes_{j=1}^L V_j$
of a chain of length $L$ then reads
\begin{equation}
\label{TRA}
T(\lambda)= Str_0[K^{+}_{0}(\lambda) R_{0L}(\lambda) \cdots R_{01}(\lambda) K^{-}_{0}(\lambda) R_{01}^{-1}(-\lambda) \cdots R_{0L}^{-1}(-\lambda)]
\end{equation}
where the symbol $Str_0$ denotes the supertrace taken over an
auxiliary five-dimensional $OSp(3|2)$ superspace.  The respective
Hamiltonian with open boundary is obtained by expanding the double-row
transfer matrix (\ref{TRA}) up to the first order in the spectral
parameter $\lambda$. For the choice of reflection matrices introduced
above the Hamiltonian can be written as
\begin{equation}
	H= \sum_{j=1}^{L-1} \left.{\frac{\partial}{\partial \lambda}{\check{R}}_{j,j+1}(\lambda)}\right|_{\lambda=0}
\label{HAM}
\end{equation}
up to an additive constant.  Recall that the $R$-matrix is constructed
by means of Baxterization from its braid limit and the
$\check{R}$-matrix can be written as a sum over projectors appearing
in the tensor product of two copies of the five-dimensional vector
representations. As has been argued before this property ensures that
the Hamiltonian (\ref{HAM}) commutes with the generators of $\ospq$
(see also our discussion of the isotropic model below) \cite{LiGo96}.

In principle it is also possible to express the Hamiltonian
(\ref{HAM}) in terms of the generators of the $OSp(3|2)$
superalgebra. To this end we recall that representations of this
superalgebra have been previously investigated by Van der Jeugt
\cite{Jeugt84}. The even part of the $OSp(3|2)$ superalgebra is
isomorphic to $SO(3) \oplus Sp(2)$ and we shall denote the
corresponding generators by the operators $\{\tau^{z},\tau^{\pm}\}$
and $\{\sigma^{z}, \sigma^{\pm}\}$, respectively.  The odd subspace of
the $OSp(3|2)$ is constituted by another six fermionic generators
which here are going to be represented by the operators
$\{c^{\pm}, d^{\pm}, f^{\pm}\}$. In terms of the $5 \times 5$ Weyl
matrices $e_{ij}$, whose entries are 1 on the $i$-th row and the
$j$-th column and zero elsewhere, the twelve generators of the
$OSp(3|2)$ superalgebra can be given explicitely in the $fbbbf$
grading. For the reader's convenience we have listed them together
with the originial notation used in Ref.~\cite{Jeugt84} in
Table~\ref{tab1}.
\begin{table}[t]
\begin{center}
\begin{tabular}{|c|c|c|c|}
  \hline
  \multicolumn{2}{|c|}{Generator} & {~~Weyl~Matrix~~} & ~~$\mathbb{Z}_2$ parity~~ \\ 
  ~~~~our notation~~~~ & notation from \cite{Jeugt84} & & \\\hline \hline
$\tau^z$ & $s^z$ & $e_{22}-e_{44}$ & $\mathrm{even}$ \\ \hline
$\tau^{+}$ & $s^+$ & $\sqrt{2}(e_{23}-e_{34})$ & $\mathrm{even}$ \\ \hline
$\tau^{-}$ & $s^-$ & $\sqrt{2}(e_{32}-e_{43})$ & $\mathrm{even}$ \\ \hline
$\sigma^{z}$ & $t^z$ & $(e_{11}-e_{55})/2$ & $\mathrm{even}$ \\ \hline
$\sigma^{+}$ & $ t^+$ & $-e_{15}$ & $\mathrm{even}$ \\ \hline
$\sigma^{-}$ & $t^-$ & $-e_{51}$ & $\mathrm{even}$ \\ \hline
$c^{+}$ & $R_{1,\frac12}$ & $e_{25}-e_{14}$ & $\mathrm{odd}$ \\ \hline
$c^{-}$ & $R_{-1,-\frac12}$ & $-e_{52}-e_{41}$ & $\mathrm{odd}$ \\ \hline
$d^{+}$ & $R_{0,-\frac12}$ & $e_{31}+e_{53}$ & $\mathrm{odd}$ \\ \hline
$d^{-}$ & $R_{0,\frac12}$ & $e_{35}-e_{13}$ & $\mathrm{odd}$ \\ \hline
$f^{+}$ & $R_{-1,\frac12}$ & $e_{12}-e_{45}$ & $\mathrm{odd}$ \\ \hline
$f^{-}$ & $R_{1,-\frac12}$ & $e_{21}+e_{54}$ & $\mathrm{odd}$ \\ \hline
\end{tabular}
\caption{The five-dimensional representation of the 
six bosonic and six fermionic generators
of the $OSp(3|2)$ superalgebra in terms 
of the Weyl matrices ordered in the $fbbbf$ grading.}
\label{tab1}
\end{center}
\end{table}
The Hamiltonian of the open spin chain (\ref{HAM}) commutes with
$U(1)$ charges of the Cartan subalgebra of $OSp(3|2)$ which are
directly related to the azimuthal bosonic generators of the $SO(3)$
and $Sp(2)$ subalgebras:
\begin{equation}
	\label{cons}
	[H, \sum_{j=1}^{L} \tau_j^{z}]=
	[H, \sum_{j=1}^{L} \sigma_j^{z}]=0\,.
\end{equation}

We further remark that the quadratic Casimir $C_{j,j+1}$ of $OSp(3|2)$
acting on pair of sites $(j,j+1)$ in terms of these generators is
\begin{eqnarray}
\label{CAS}
C_{j,j+1} &=& \frac{\left(\tau_j^{+} \tau_{j+1}^{-} +\tau_j^{-} \tau_{j+1}^{+}\right)}{2} +\tau_j^z \tau_{j+1}^z -2\left( \sigma_{j}^{+} \sigma_{j+1}^{-} + \sigma_j^{+} \sigma_{j+1}^{-} \right) 
-4\sigma_{j}^z \sigma_{j+1}^{z} \nonumber \\
&+&\left(c_{j}^{+} c_{j+1}^{-}-c_{j}^{-} c_{j+1}^{+} \right)
+ \left( d_{j}^{+} d_{j+1}^{-}-d_{j}^{-} d_{j+1}^{+} \right)
+\left( f_{j}^{+} f_{j+1}^{-}-f_{j}^{-} f_{j+1}^{+} \right)
\end{eqnarray}
where the tensor products among the fermionic degrees of freedom have
to be understood in the graded sense.  The degeneracies of the
eigenvalues of this Casimir operator are compatible with expected
Clebsch-Gordon decomposition
$(0;\frac12)\otimes(0;\frac12) = (0;0) \oplus (0;1) \oplus
(1;\frac12)$, corresponding to the identity, the $12$-dimensional
adjoint and a non-fundamental representation \cite{Jeugt84}.  At this
point we have the basic ingredients to represent the quantum group
Hamiltonian (\ref{HAM}) in terms of the generators of the superalgebra
$OSp(3|2)$. The final expression for the Hamiltonian is quite
cumbersome and is given in Appendix~\ref{app:HAMq}.

In the limit $\gamma \rightarrow 0$ the Hamiltonian (\ref{HAM}) can be
written in terms of the nearest-neighbour Casimir operator,
\begin{equation}
H= -\sum_{j=1}^{L-1} \left(C_{j,j+1}-3C_{j,j+1}^2\right)  -4(L-1)\,,
\end{equation}
making explicit the the $OSp(3|2)$ invariance at the isotropic point.

\section{The Bethe ansatz solution}
The Hamiltonian of the $\defq$-deformed $OSp(3|2)$ superspin chain
with \emph{periodic} boundary condition has been diagonalized using
the algebraic Bethe ansatz in Ref.~\cite{GaMa04}.  Within the
framework of the analytical Bethe ansatz it has been found that the
Bethe equations for the corresponding \emph{open} spin chain with
quantum algebra invariance can be obtained from those for the periodic
model based on the so-called 'doubling postulate'
\cite{MeNe92a,ArMN95a,ArMN95,YuBa95c}.  We have checked that the
postulate applies for the $\ospq$ model by comparing the eigenenergies
obtained by exact diagonalization of the Hamiltonian (\ref{HAM}) with
those obtained solving the doubled Bethe equations given below for
several low-lying states up to $L=8$.

With these Bethe equations the spectrum of the model is built starting
from suitable highest weight reference states.  For models based on
superalgebras it is well known that the explicit form of the Bethe
equations depend on the choice of the grading,
see. e.g. \cite{EsKo92}.  For the $\ospq$ superspin chain this amounts
to two different formulations of the Bethe ansatz, i.e.\ in the
grading $fbbbf$ starting from a reference state in the
$(L-1;\frac12)$-multiplet and in the grading $bfbfb$ starting from the
$(0;L/2)$-multiplet \cite{FrMa15}.
In what follows we shall discuss the form of the Bethe equations for
these two possible gradings.

\subsection{The fbbbf grading}
Applying the doubling procedure to the $fbbbf$ Bethe equations of the
periodic $\ospq$ model \cite{GaMa04,FrHM19} we find that the
eigenstates of the open $\ospq$ invariant superspin chain (\ref{HAM})
are parametrized by solutions to the following set of Bethe equations,
\begin{equation}
  \label{bethefbbbf}
  \begin{aligned}
    &\left[f_{1/2}\left(\lambda_{j}^{(1)}\right)\right]^{{2L}}=
    \prod_{k=1}^{{L-n_1-n_2}}
    f_{1/2}\left({\lambda_{j}^{(1)}-\lambda_{k}^{(2)}}\right)\,
    f_{1/2}\left(\lambda_{j}^{(1)}+\lambda_{k}^{(2)}\right)\,,\quad j=1,\cdots,{L-n_1} , 
    \\
    &\prod_{k=1}^{{L-n_1}}
    f_{1/2}\left(\lambda_{j}^{(2)}-\lambda_{k}^{(1)}\right)\,
    f_{1/2}\left(\lambda_{j}^{(2)}+\lambda_{k}^{(1)}\right)=
    \\ &\qquad= 
	  \prod_{\stackrel{k=1}{k \ne j} }^{{L-n_1-n_2}}
    f_{1/2}\left(\lambda_{j}^{(2)}-\lambda_{k}^{(2)}\right)\,
    f_{1/2}\left(\lambda_{j}^{(2)}+\lambda_{k}^{(2)}\right)\,, 
    \quad j=1,\cdots,{L-n_1-n_2} \,,
  \end{aligned}
\end{equation}
where the function $f_s(\lambda)$ is defined as
\begin{equation}
f_s(\lambda)= \frac{\sinh(\lambda +is\gamma)} 
{\sinh(\lambda -is\gamma)}\,.
\end{equation}
In (\ref{bethefbbbf}) $n_1$, $n_2$ are the eigenvalues of the $U(1)$
charges (\ref{cons}) for a highest weight state in the multiplet
$(\repp;\repq)=(n_1-1;\frac12(n_2+1))$.  Its energy is given in terms
of the Bethe roots from the first level as,
\begin{equation}
\label{enefbbbf}
  E(\{\lambda_j^{(a)}\},L)= \sum^{L-n_1}_{j=1}
    \frac{2 \sin\gamma}{\cos\gamma-\cosh(2\lambda_j^{(1)})}\,.
\end{equation}

In order to study the thermodynamic limit properties we first
diagonalized the Hamiltonian (\ref{HAM}) for lattice sizes $L \leq
8$. We next solve numerically the Bethe equations (\ref{bethefbbbf})
for some $U(1)$-sectors $(n_1,n_2)$ and compare the eigenenergies
(\ref{enefbbbf}) with the spectrum obtained by exact diagonalization
of Hamiltonian (\ref{HAM}). For low-lying energy states, similar as in
the periodic case \cite{FrHM19}, we find that the Bethe root
configurations on both levels are dominated by two-strings with
$\mathrm{Re}\left(\lambda_j^{(1,2)}\right)\geq0$ and
$\mathrm{Im}\left(\lambda_j^{(1,2)}\right) \simeq \pm \gamma/4$.  As
the system size grows the difference among the root configurations of
the two levels becomes exponentially close. Therefore, for
$L \rightarrow \infty$ the respective string hypothesis may be
formulated as
\begin{equation}
  \label{stringfbbbf}
  \lambda^{(1)}_{j}= \xi_j \pm i\frac{\gamma}{4}\,,\quad
  \lambda^{(2)}_{j}= \xi_j \pm i\frac{\gamma}{4}\,,\quad
	\xi_j \in \mathbb{R}^{+}\,. 
\end{equation}
In the thermodynamic limit these strings fill the positive part of the
real axis and the root configuration can be described in terms of
their density $\sigma_{L}(\xi)$ within the root density approach
\cite{YaYa69}.  Symmetrizing the density around the origin we obtain
the following linear integral equation for $\sigma_L(\xi)$:
\begin{equation}
\label{eq:densityfbbbf}
\begin{aligned}
2\pi \sigma_{L}(\xi) &+ \int_{-\infty}^{+\infty} \mathrm{d} \xi^{'}
	\left[2\Psi\left(\xi-\xi^{'},\frac{\gamma}{2}\right)
	+\Psi\left(\xi-\xi^{'},\gamma\right) \right] \sigma_{L}(\xi^{'}) \\
	&= 2\left[\Psi\left(\xi,\frac{3\gamma}{4}\right)
	+\Psi\left(\xi,\frac{\gamma}{4}\right) \right]+
 \frac{1}{L} 
	\left[
	2\Psi\left(\xi,\frac{\gamma}{2}\right)
	+\Psi\left(\xi,\gamma\right)-2\Psi\left(2\xi,\gamma\right) \right]\,,
\end{aligned}
\end{equation}
where
$\Psi(\xi,\gamma)=\frac{2\sin(2\gamma)}{\cosh(2x)-\cos(2\gamma)}$.
This equation can be solved by Fourier transformation order by order
in $L$.  To leading order one finds
\begin{equation}
\label{sigma}
	\sigma_{\infty}(x)= \frac{2}{\gamma \cosh(2\pi x/\gamma)}\,,
\end{equation}
from which we reproduce the ground state energy density
\begin{equation}
  \varepsilon_{\infty} =-2\cot{\frac{\gamma}{2}}\,.
\end{equation}
The low energy excitations above the ground state are gapless with a
linear dispersion relation $\epsilon(p) \simeq v_F |p|$ where the
Fermi velocity is $v_F= 2\pi/\gamma$. These quantities are already
known from the $\defq$-deformed model with periodic boundary
conditions \cite{FrHM19}.  Similarly, one obtains the surface energy
$f_\infty$ from the $\mathcal{O}(L^{-1})$ contribution to the solution
of (\ref{eq:densityfbbbf}).  After some simplifications we find
\begin{equation}
	f_{\infty}^{(OSp)}=
	2\cot\frac{\gamma}{2}\,,
\end{equation}
which is just the value for the ground state energy per site with the
opposite sign.

\subsection{The grading bfbfb }

Alternatively we can use the doubling procedure in the Bethe ansatz
solution of the model with periodic boundaries for the grading $bfbfb$
\cite{FrHM19}. In this case the spectrum of the open $\ospq$ invariant
superspin chain is parametrized by solutions of a different set of
Bethe equations,
\begin{equation}
  \label{bethebfbfb}
  \begin{aligned}
      &\left[f_{1/2}\left(\lambda_{j}^{(1)}\right)\right]^{{2L}}=
	  \prod_{k=1}^{{L-m_1-m_2}}
    f_{1/2}\left({\lambda_{j}^{(1)}-\lambda_{k}^{(2)}}\right)\,
    f_{1/2}\left(\lambda_{j}^{(1)}+\lambda_{k}^{(2)}\right)\,,
	  \quad j=1,\cdots,{L-m_2}, 
    \\
	  &\prod_{k=1}^{{L-m_2}}
    f_{1/2}\left(\lambda_{j}^{(2)}-\lambda_{k}^{(1)}\right)\,
    f_{1/2}\left(\lambda_{j}^{(2)}+\lambda_{k}^{(1)}\right) =
    \\ &\qquad = 
	  \prod_{\stackrel{k=1}{k \ne j} }^{{L-m_1-m_2}}
    f_{-1/2}\left(\lambda_{j}^{(2)}-\lambda_{k}^{(2)}\right) \,
    f_{-1/2}\left(\lambda_{j}^{(2)}+\lambda_{k}^{(2)}\right) \
    f_1\left(\lambda_{j}^{(2)}-\lambda_{k}^{(2)}\right)\,
    f_1\left(\lambda_{j}^{(2)}+\lambda_{k}^{(2)}\right)\,,\\ 
	  & \qquad\qquad j=1,\cdots,{L-m_1-m_2} \,,
  \end{aligned}
\end{equation}
and the energy of the Hamiltonian (\ref{HAM}) corresponding to a
particular root configuration is
\begin{equation}
\label{eneBFBFB}
  E\left(\{\lambda_j^{(a)}\},L\right)
  = -\sum^{L-m_2}_{j=1} \frac{2 \sin\gamma}
	{\cos\gamma-\cosh(2\lambda_j^{(1)})} -2(L-1) \cot\gamma\,.
\end{equation}
As mentioned above, the Bethe states for $fbbbf$ and $bfbfb$ are
constructed starting from different reference states.  Solutions to
(\ref{bethebfbfb}) parametrize the state with $U(1)$-charges
$(n_1,n_2)=(m_1+1,m_2-1)$ in a $(\repp;\repq)=(m_1;{m_2}/2)$
multiplet.

It turns out that the thermodynamic limit in the $bfbfb$ grading
simplifies studies in the subsector with quantum number $m_1=0$
because the number of Bethe roots in both levels are the same.
Numerical solution of the Bethe equations (\ref{bethebfbfb}) for small
systems tell us that the Bethe root configurations are dominated by
pairs of complex rapidities with positive real parts and
$\mathrm{Im}\left(\lambda_j^{(1)}\right) \simeq \pm 3\gamma/4$,
$\mathrm{Im}\left(\lambda_j^{(2)}\right) \simeq \pm \gamma/4$.  Once
again the difference of their real parts become exponentially small
for large $L$. Similar to what has been found for the periodic case
\cite{FrHM19}, the string hypothesis for the $bfbfb$ grading is given
by,
\begin{equation}
\label{stringbfbfb}
  \lambda^{(1)}_{j,\pm} = \xi_j \pm i\frac{3\gamma}{4}\,,\quad
  \lambda^{(2)}_{j,\pm} = \xi_j \pm i\frac{\gamma}{4}\,,\quad
	\xi_j\in\mathbb{R}^{+}\,.
\end{equation}

The root density approach based on this string assumption can be used
to reproduce the bulk energy density, Fermi velocity of low energy
excitations and surface energy obtained in the grading $fbbbf$.

\section{The $\xxzq$ spin $S=1$ $XXZ$ model}
\label{sec:xxzq}
The Hamiltonian of the integrable $S=1$ $XXZ$ model with quantum 
algebra symmetry is
\cite{PaSa90,BMNR90}
\begin{equation}
	\label{HAMXXZ}
H = \sum_{j=1}^{L-1} H_{j,j+1} + i(S_L^z-S_1^ z)
\end{equation}
where the nearest-neighbour bulk terms $H_{j,j+1}$ are those of the
integrable spin-1 model introduced by Zamolodchikov and Fateev
\cite{ZaFa80},
\begin{equation}
\begin{aligned}
	H_{j,j+1} =& \frac{1}{\sin(\gamma)} \left[{\bf S}_j\cdot {\bf S}_{j+1}-({\bf S}_j\cdot {\bf S}_{j+1})^2 \right] 
	-\tan(\frac{\gamma}{2}) \left[ S_j^z S_{j+1}^z +(S_j^ z)^ 2+(S_{j+1}^z)^2-(S_j^z S_{j+1}^z)^ 2 \right]\\
	&+ \frac{4\sin^2(\frac{\gamma}{4})}{\sin(\gamma)} \left[ (S_j^{x} S_{j+1}^{x} + S_j^{y} S_{j+1}^{y}) S_j^{z} S_{j+1}^z 
	+S_j^z S_{j+1}^z(S_j^{x} S_{j+1}^{x} + S_j^{y} S_{j+1}^{y}) \right]  \\
	&+ \left(1+2\tan(\frac{\gamma}{2})\right)\,\mathbf{1}_j \mathbf{1}_{j+1} \,,
\end{aligned}
\end{equation}
where the last constant term is a convenient normalization for the
spectral relationship to the $\ospq$ superspin chain. The
${\bf S} =(S^x,S^y,S^z)$ are the spin-1 generators of $SU(2)$
\begin{equation}
S^x=\frac1{\sqrt{2}}\left( \begin{array}{ccc} 
	0 &  1 & 0   \\
	1 &  0 & 1   \\
	0 &  1 & 0  \\
	\end{array}
	\right)\,,\quad
S^y=\frac1{\sqrt{2}}\left( \begin{array}{ccc} 
	0 &  -i & 0   \\
	i &  0 & -i   \\
	0 &  i & 0  \\
	\end{array}
	\right)\,,\quad
S^z=\left( \begin{array}{ccc} 
	1 &  0 & 0   \\
	0 &  0 & 0   \\
	0 &  0 & -1  \\
	\end{array}
	\right)
\end{equation}
and $\mathbf{1}$ is the $3\times3$ identity.  The Hamiltonian
(\ref{HAMXXZ}) has been diagonalized using the algebraic Bethe ansatz
\cite{MeNR90}.  The eigenstates of this model are parameterized by the
$L-n$ complex roots $\lambda_k$ of the Bethe equations,
\begin{equation}
  \label{betheXXZ}
    \left[f_{1/2}\left(\lambda_{k}\right)\right]^{{2L}}=
	\prod_{\stackrel{\ell \neq k}{\ell=1}}^{{L-n}}
    f_{1/2}\left({\lambda_{k}-\lambda_{\ell}}\right)\,
    f_{1/2}\left(\lambda_{k}+\lambda_{\ell}\right)\,,\quad k=1,\cdots,{L-n}\, . 
\end{equation}
where $n$ is the $U(1)$-charge $\sum_{j=1}^{L}S_j^{z}$ for the highest
weight state in a $\xxzq$ spin-$j$ multiplet, i.e.\ $n=j$.  In terms
of these roots the corresponding energies of the Hamiltonian
(\ref{HAMXXZ}) are
\begin{equation}
\label{eneXXZ}
  E(\{\lambda_k\},L)= \sum^{L-n}_{k=1}
    \frac{2 \sin\gamma}{\cos\gamma-\cosh(2\lambda_k)}\,.
\end{equation}

For large $L$, the Bethe roots configurations for the low-lying
energies are essentially dominated by pairs of complex complex
conjugate roots
\begin{equation}
  \label{stringXXZ}
  \lambda_{k,\pm} \simeq \xi_k \pm i\frac{\gamma}{4}\,,\quad
	\xi_k \in \mathbb{R}^{+}\,. 
\end{equation}
In the thermodynamic limit the density of these $2$-strings is given
by the integral equation
\begin{equation}
\label{eq:densityxxz1}
\begin{aligned}
2\pi \sigma_{L}(\xi) &+ \int_{-\infty}^{+\infty} d \xi^{'}
	\left[2\Psi\left(\xi-\xi^{'},\frac{\gamma}{2}\right)
	+\Psi\left(\xi-\xi^{'},\gamma\right) \right] \sigma_{L}(\xi^{'}) \\
	&= 2\left[\Psi\left(\xi,\frac{3\gamma}{4}\right)
	+\Psi\left(\xi,\frac{\gamma}{4}\right) \right]+
 \frac{1}{L} 
	\left[2\Psi\left(\xi,\frac{\gamma}{2}\right)
	+\Psi\left(\xi,\gamma\right) +2\Psi\left(2\xi,\gamma\right)\right]
\end{aligned}
\end{equation}
obtained within the  root density approach.

Comparing Eq.~(\ref{eq:densityxxz1}) with the corresponding result for
the $\ospq$ superspin chain given by Eq.~(\ref{eq:densityfbbbf}) we
note that they only differ in the sign of the driving term
$2\Psi(2\xi,\gamma)/L$.  Therefore both the bulk energy density
$\varepsilon_\infty$ and the Fermi velocity $v_F$ of the models
(\ref{HAM}) and (\ref{HAMXXZ}) coincide in the critical region and
only the corresponding surface energies $f_\infty$ differ. Considering
the thermodynamic limit of the relation (\ref{eq:densityxxz1}) we find
\begin{equation}
	f_{\infty}^{(XXZ)}= 2\int_{-\infty}^{+\infty}\mathrm{d}x\, \frac{\cosh[(\pi-\gamma)x] \tanh(\gamma x)}{\sinh(\pi x)}
\end{equation}
for the integrable quantum group invariant $\xxzq$ spin-$1$ chain.

\begin{figure}[t]
\begin{center}
  \includegraphics[width=0.45\textwidth]{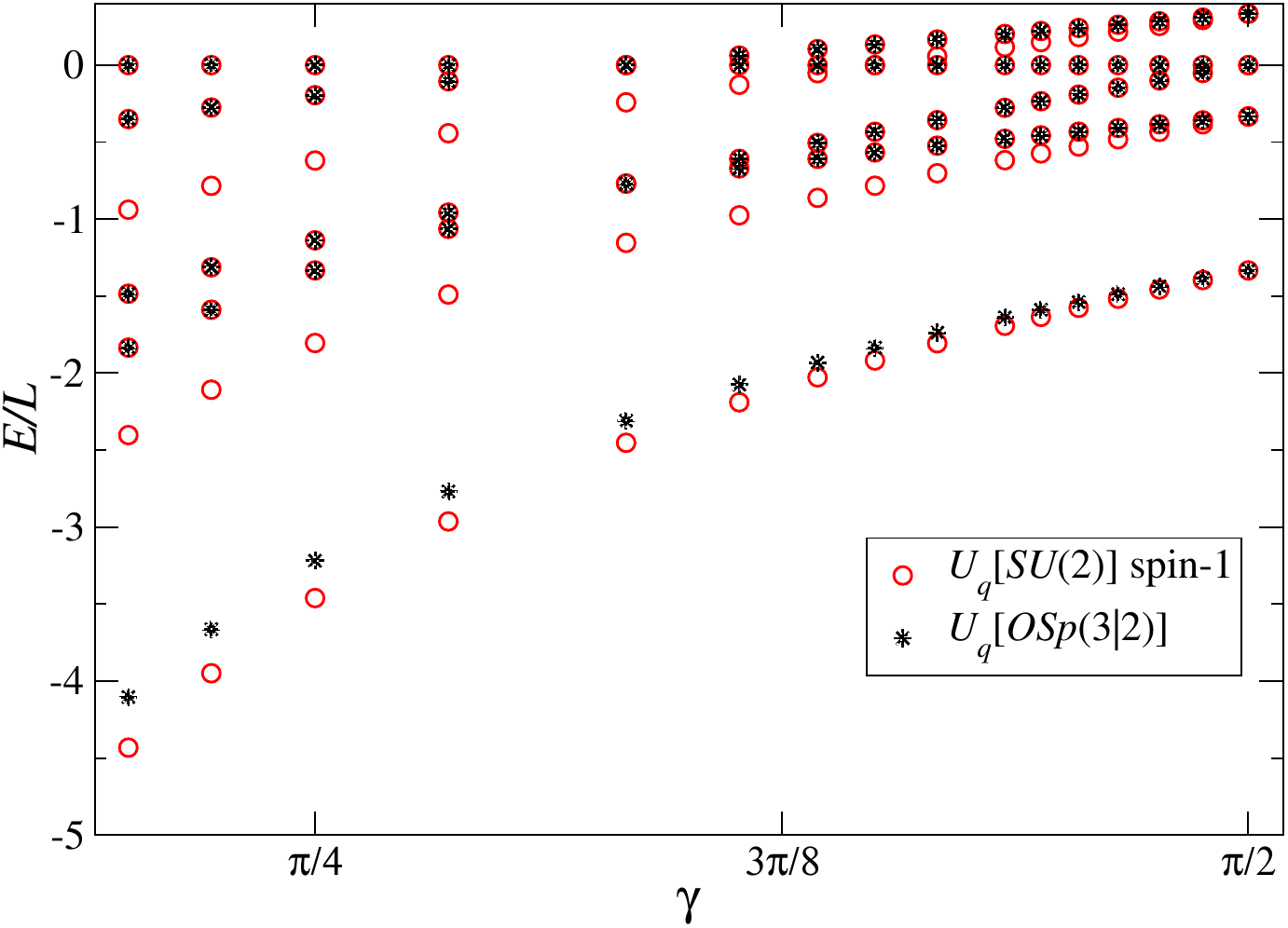}\hspace*{2mm}
  \includegraphics[width=0.45\textwidth]{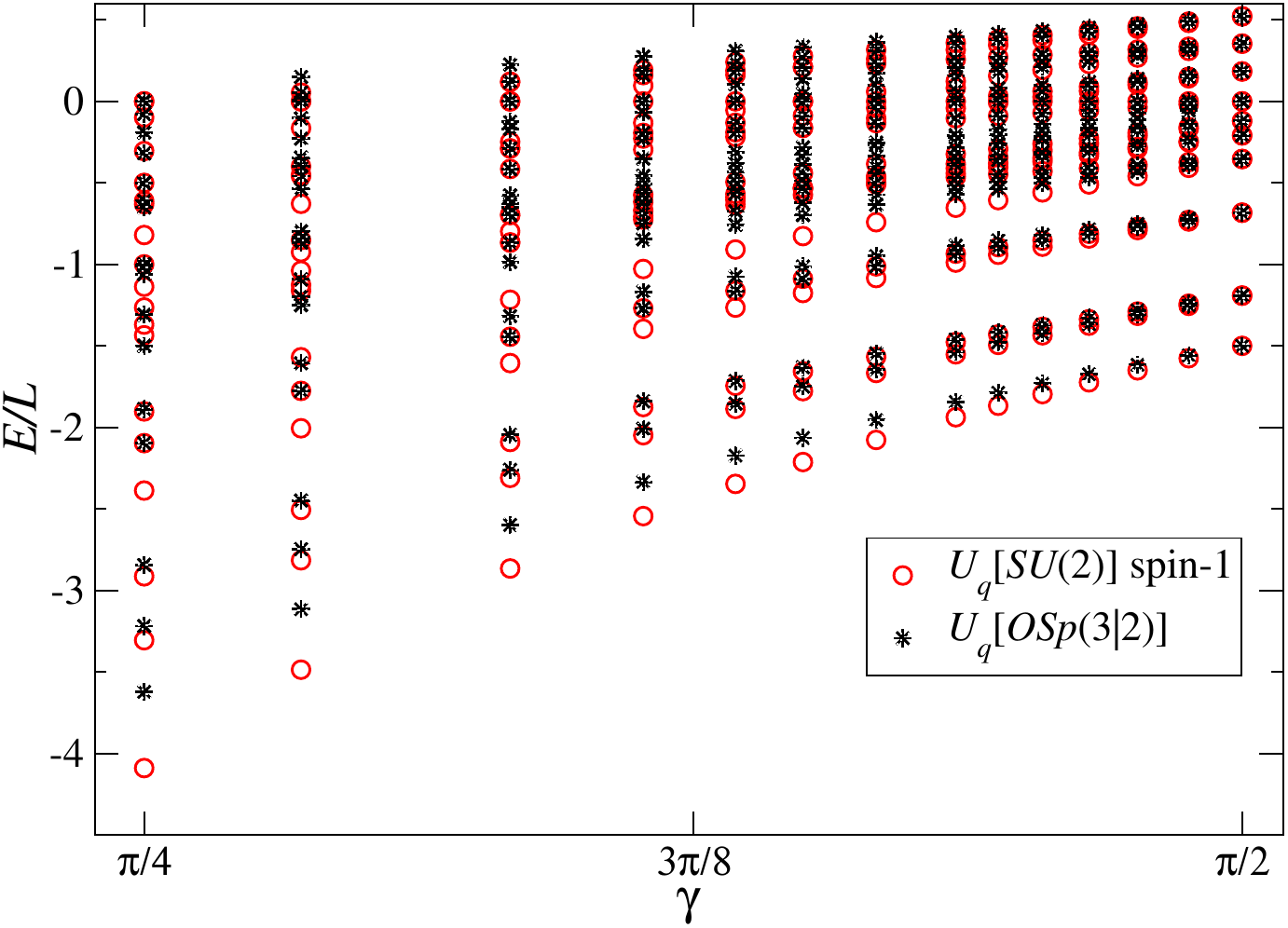}
  \end{center}
  \caption{\label{fig:comp2XXZ}Spectra of the $\ospq$ superspin chain
    and the quantum group invariant $XXZ$ spin-$1$ model as a function
    of the anisotropy $\gamma$ for $L=3$ (left panel) and $L=4$ (right
    panel). Note that for $\gamma=\pi/2$ the spectra coincide up to
    degeneracies.}
\end{figure}

Note, however, that $\Psi(2\xi,\gamma)$ vanishes for $\gamma=\pi/2$
and therefore both the bulk and surface energies of the two models
coincide. It has already been observed that the $XXZ$ spin-$1$ chain
and the $\defq$-deformed $OSp(3|2)$ model with periodic boundary
conditions have some common eigenvalues for this value of the
anisotropy\cite{FrHM19}.  Motivated by that we have compared the
eigenspectra of the quantum group invariant Hamiltonians (\ref{HAM})
and (\ref{HAMXXZ}) for small lattice sizes using exact
diagonalizations. Remarkably, we find perfect matching of the spectra
of these Hamiltonians for $\gamma=\frac{\pi}{2}$ apart from
degeneracies due to the different sizes of the Hilbert space, see
Figure~\ref{fig:comp2XXZ}.
Within our numerical precision this feature has also been checked for
the complete spectrum of both models up to the size $L=8$.  In
addition we have observed this correspondence of the spectra continues
to hold from the point of view of solutions of the corresponding Bethe
ansatz equations: for small $L$ we find that the level-$1$ Bethe roots
for the state with $(\repp;\repq)=(0;1)$ in the $fbbbf$ grading
approach those for the $n=1$ state in the $\xxzq$ chain as
$\gamma\to\pi/2$.  As a consequence of (\ref{enefbbbf}) and
(\ref{eneXXZ}) their energies coincide in this limit.

\paragraph*{The critical behaviour.}
The low energy spectrum of the integrable spin-$1$ $XXZ$ spin chain in
the critical regime $0 \leq \gamma \leq {\pi}$ subject to various
boundary conditions has been studied extensively
\cite{FrSZ88,AlMa89,AlMa90,FrYF90,Mart90,KoSa93}.  The finite-size
spectrum of the quantum group invariant model has first been
investigated for the lowest states with $\xxzq$-spin $j=0$ and $1$ in
Ref.~\cite{Mart90} for even length chains giving the conformal anomaly
and the conformal weight $h_{j=1}$ as a function of the anisotropy
\begin{equation}
  \label{eq:xxzq_c}
	c_{\text{qg}}= \frac{3}{2} \left( 1- \frac{2\gamma^2}{\pi (\pi-\gamma)} \right)\,, \quad
	h_{j=1} = 1-\frac{\gamma}{\pi}\,.
\end{equation}
Later on it has been proposed that the full operator content of the
$\xxzq$-invariant spin chain can be obtained by combining a twisted
free boson field with compactification radius depending on the
anisotropy and an Ising field with free boundaries.  More precisely
the conformal weights, in our notation, are \cite{KoSa93}
\begin{equation}
   \label{eq:xxzq_KoSa}
    h_j = \frac{j}{2}\left(j-\frac\gamma\pi\left(j+1\right)\right) +
       h_I\,,
\end{equation}
where $h_I=0$ or $\frac12$ depending on whether $j$ is even or odd is
related to the Ising degree of freedom \cite{GeRi86a}.  Note that
$c_{\text{qg}}$ vanishes for $\gamma=\frac{\pi}{2}$ together with the
subleading corrections to scaling to the ground state. This is
accordance with the mentioned correspondence with the $\ospq$
symmetric superspin chain.

\begin{figure}[t]
\begin{center}
  \includegraphics[width=0.45\textwidth]{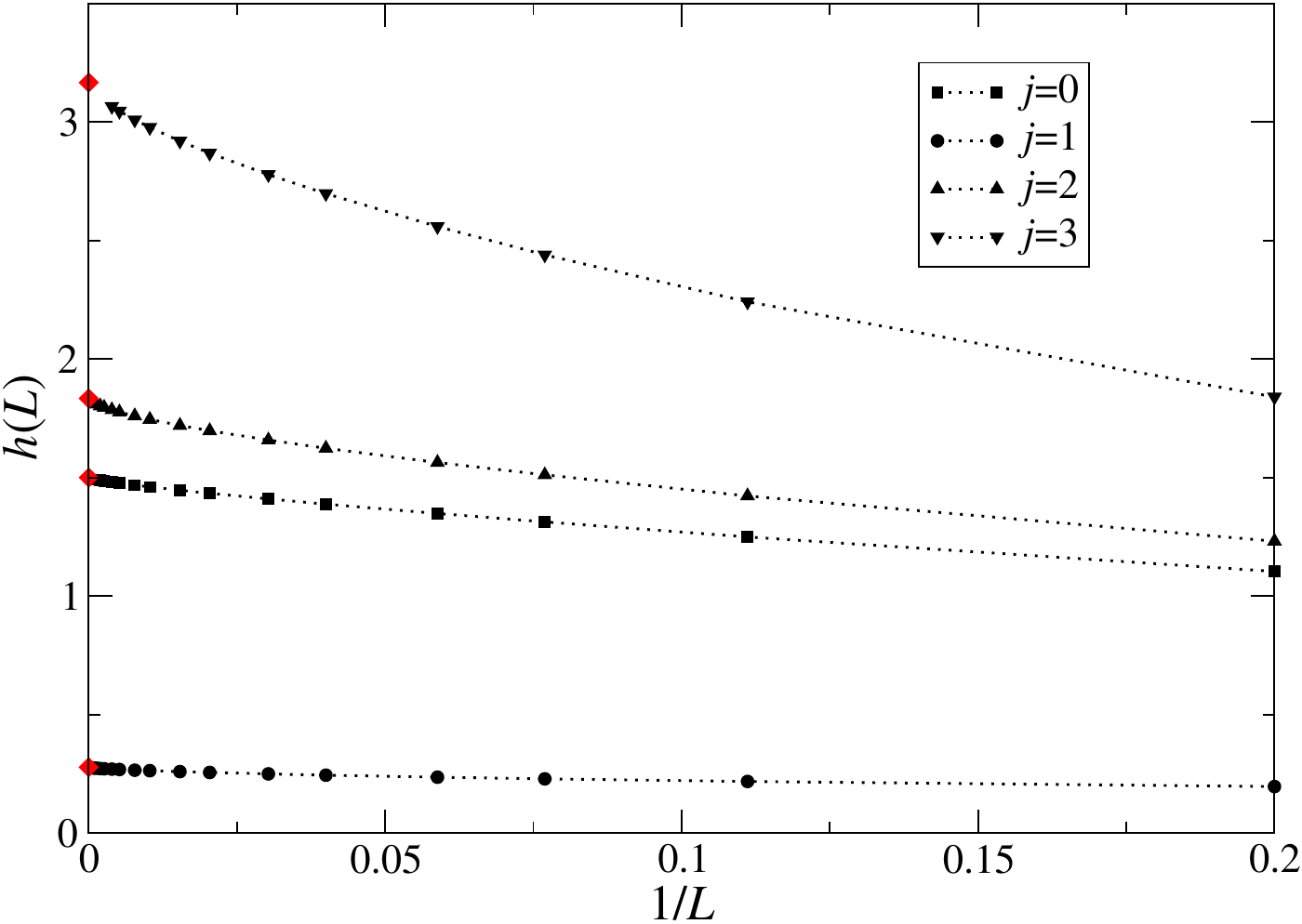}
  \end{center}
  \caption{\label{fig:xxz1_odd} Finite-size estimates
    $h_j(L)=(L/\pi v_F) \left(E_{j}(L)-E_0(L)\right)$ of the conformal
    weights corresponding to the ground states of the quantum group
    invariant spin-$1$ chain (\ref{HAMXXZ}) in the sectors with total
    spin $j=0,1,2,3$ for $\gamma=2\pi/9$ and odd length $L$. Red
    diamonds indicate the values proposed in (\ref{eq:xxzq_KoSa}) with
    (\ref{eq:xxzq_Ising}).}
\end{figure}

\begin{figure}[t]
\begin{center}
  \includegraphics[width=0.45\textwidth]{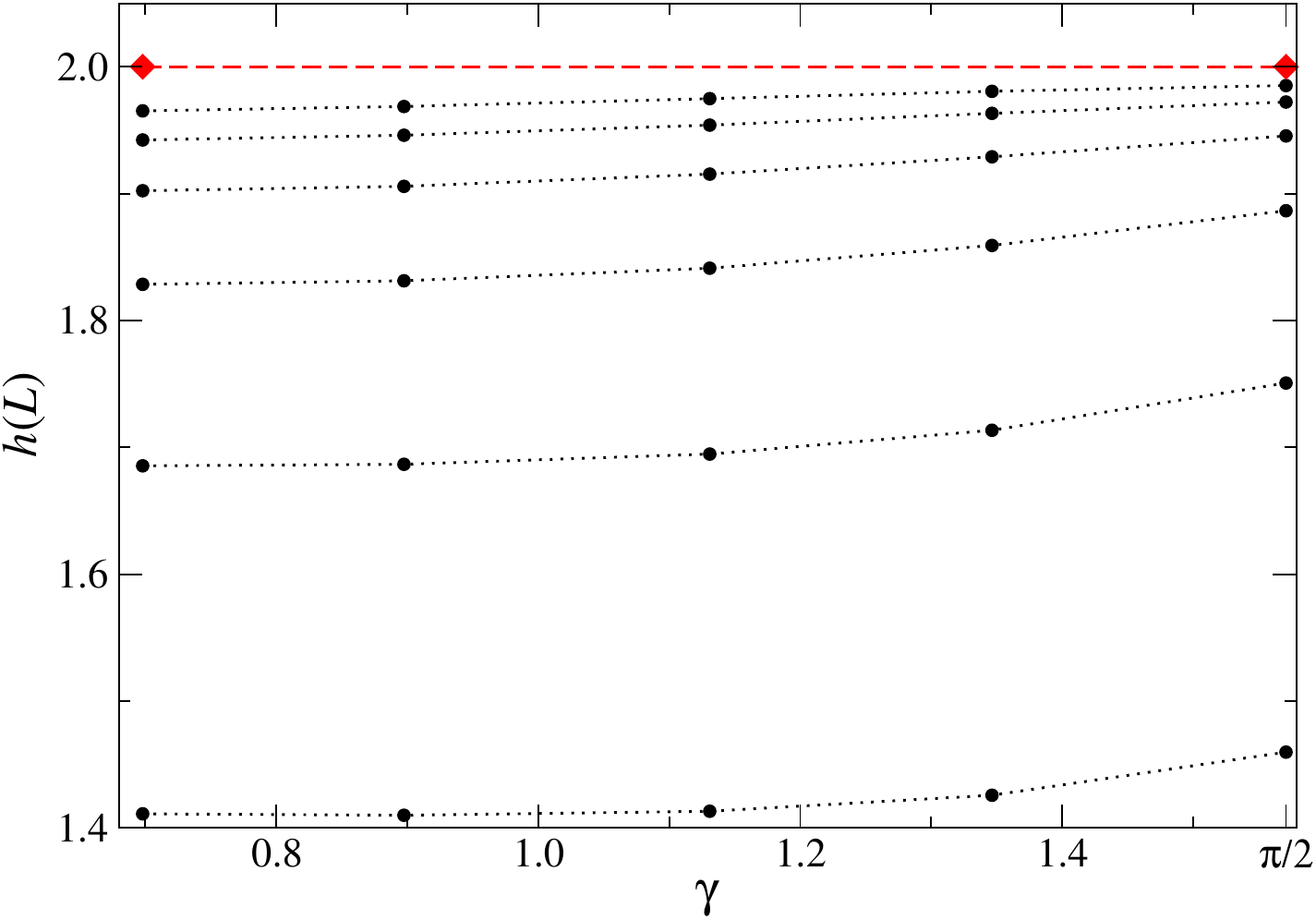}
  \end{center}
  \caption{\label{fig:xxz1_stress} Finite-size estimates
    $h(L)=(L/\pi v_F) \left(E_{0,x}(L)-E_0(L)\right)$ of the conformal
    weight corresponding to the first excitation in the total spin
    $j=0$ sector of the quantum group invariant spin-$1$ chain
    (\ref{HAMXXZ}) for various $\gamma$.  The dotted lines connecting
    data for even lengths $L=2^k$, $k=2,\dots,7$ are a guide to the
    eye only. The data converge to $h\equiv2$ independent of $\gamma$
    as $L\to\infty$ (red dashed line).}
\end{figure}

As a consequence of the antiferromagnetic character of the quantum
group invariant spin chain the ground state is typically frustrated
which may lead to an excited state with distinct exponent when the
thermodynamic limit is taken for $L$ odd.  Based on our numerical
results shown in Fig.~\ref{fig:xxz1_odd}
we are led to complement the proposal of Ref.~\cite{KoSa93} for the
operator content of the conformal field theory describing the
continuum limit of the $\xxzq$ spin-$1$ chain to be given by
(\ref{eq:xxzq_KoSa} but with
\begin{equation}
\label{eq:xxzq_Ising}
    h_I = \begin{cases}
      0       & \text{for~} L+j \text{~even}\\
      \frac12 & \text{for~} L+j \text{~odd}
      \end{cases}\,.
\end{equation}
In addition we have identified the Bethe root configuration for the
first excitation in the $j=0$ sector for chains with even $L$. This is
a descendent of the identity with conformal weight $h=2$ independent
of the anisotropy which we identify with the stress tensor.  The
results are exhibited Figure~\ref{fig:xxz1_stress}.

We emphasize that this spectrum of conformal weights is discrete for
all values of $\gamma$.  At rational values of $\gamma/\pi$ the
conformal weights can be rearranged in terms of an extended algebra
with a finite number of primaries.  Specifically, for $\gamma=\pi/2$
(\ref{eq:xxzq_KoSa}) with (\ref{eq:xxzq_Ising}) take only integer and
half-odd integer values.

Given the spectral correspondence with the $\ospq$ symmetric superspin
chain the critical properties of the quantum group invariant spin-$1$
chain will provide additional input to elaborate on the conformal
content of the $\ospq$ symmetric superspin chain below.

\section{Finite-size spectrum  of the $\ospq$ chain}
Based on the exact diagonalization of small systems we find that the
ground state of the $\ospq$ superspin chain with quantum-group
invariant boundary conditions is a $(0;0)$-singlet for $L$ even
($(0;1/2)$-quintet for $L$ odd) with energy
\begin{equation}
\label{eq:energy0}
    E_0 \equiv L\epsilon_\infty + f_\infty^{(OSp)} = -2(L-1)\cot\left(\gamma/2\right)\,,
\end{equation}
i.e.\ without any finite-size corrections implying that the effective
central charge of the field theory is
$c_{\text{eff}}=0$.\footnote{This has also been found for the
  isotropic $OSp(3|2)$ superspin chain with both periodic and free
  boundary conditions \cite{FrMa15,FrMa22}. Note that
  (\ref{eq:energy0}) is \emph{not} the lowest energy in the
  $\defq$-deformed $OSp(3|2)$ model subject to periodic boundary
  conditions \cite{FrHM19}.}
For odd $L$ Eq.~(\ref{eq:energy0}) can be verified by solving the
$bfbfb$ Bethe equations (\ref{bethebfbfb}) for the quintet where we
find root configurations containing $(L-1)/2$ complex conjugate pairs
of roots on each level which are arranged in groups similar to
(\ref{stringbfbfb}).  The root configurations for even $L$ contain
degenerate roots.

\begin{figure}[t]
\begin{center}
  \includegraphics[width=0.44\textwidth]{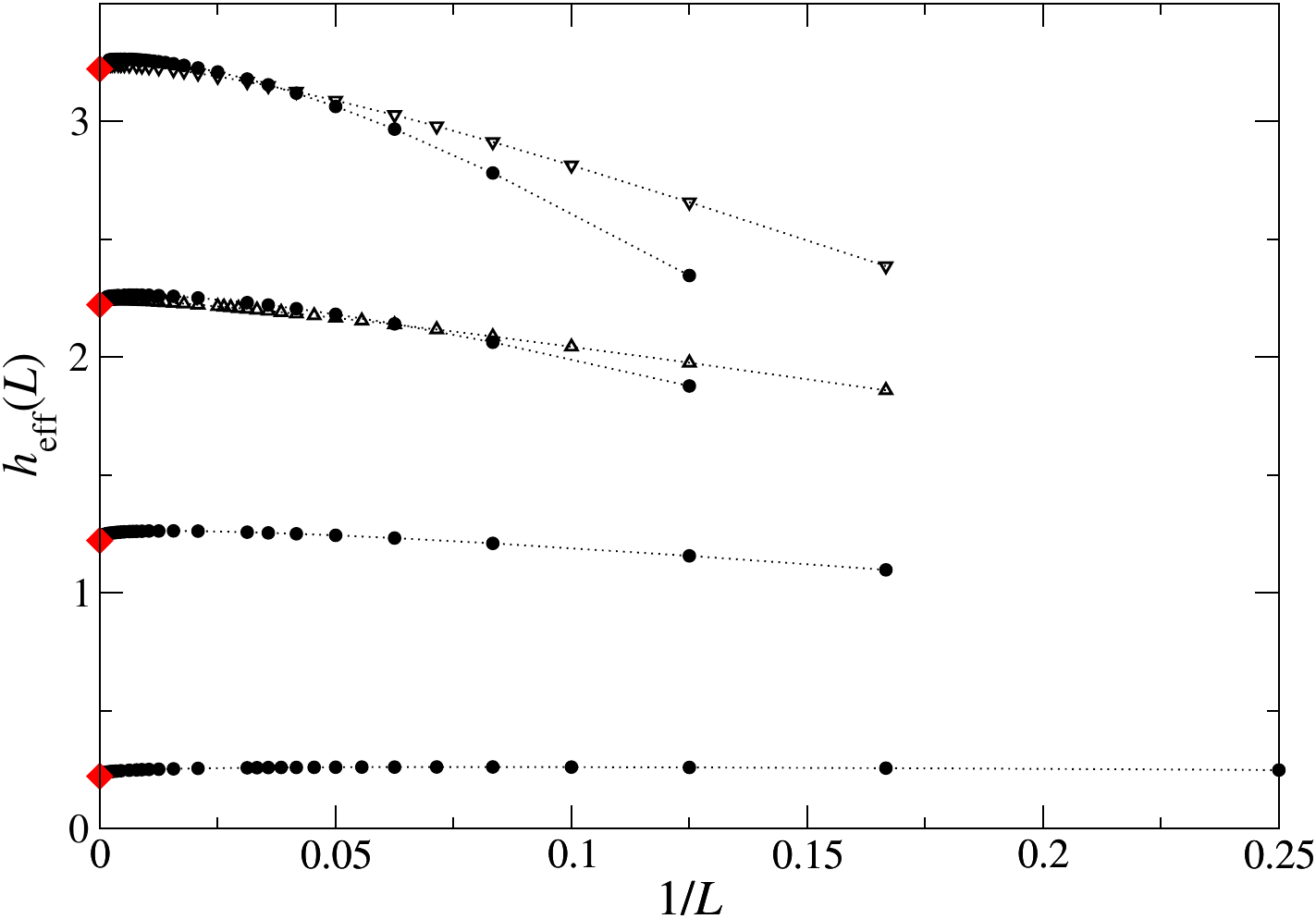}
  \hspace*{5mm}\includegraphics[width=0.44\textwidth]{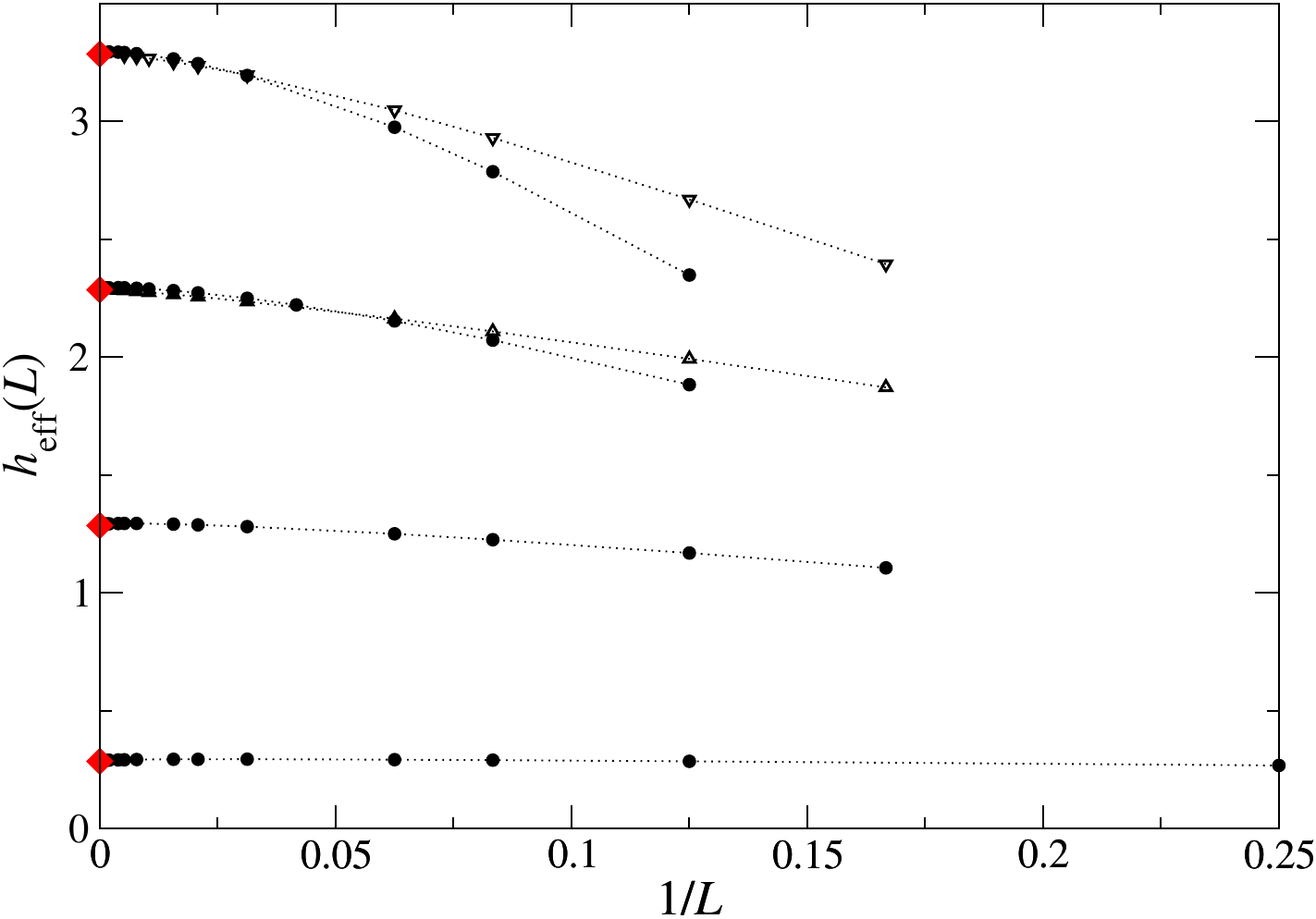}
  \vspace*{2mm}
  
  \includegraphics[width=0.45\textwidth]{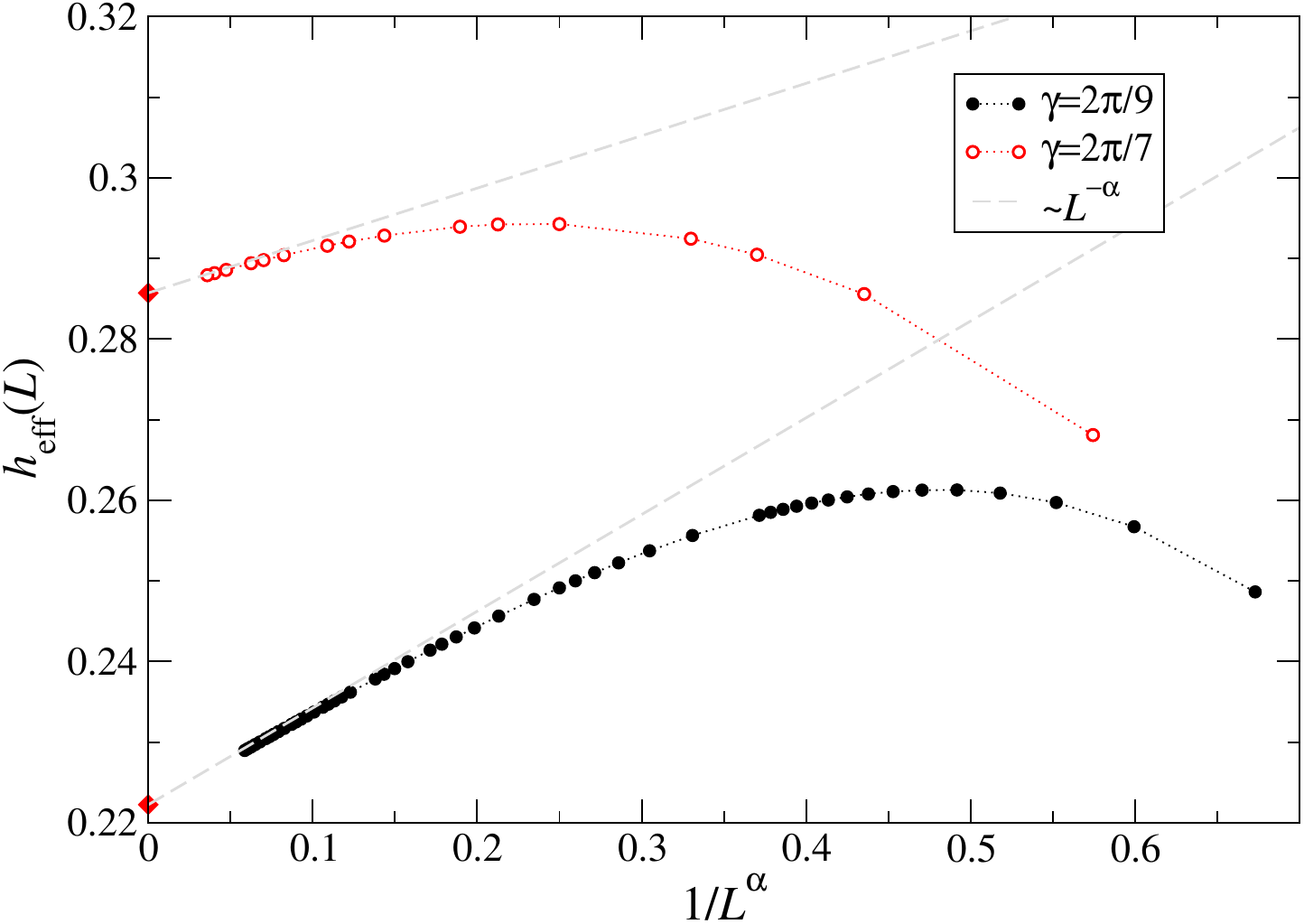}
  \hspace*{5mm}\includegraphics[width=0.45\textwidth]{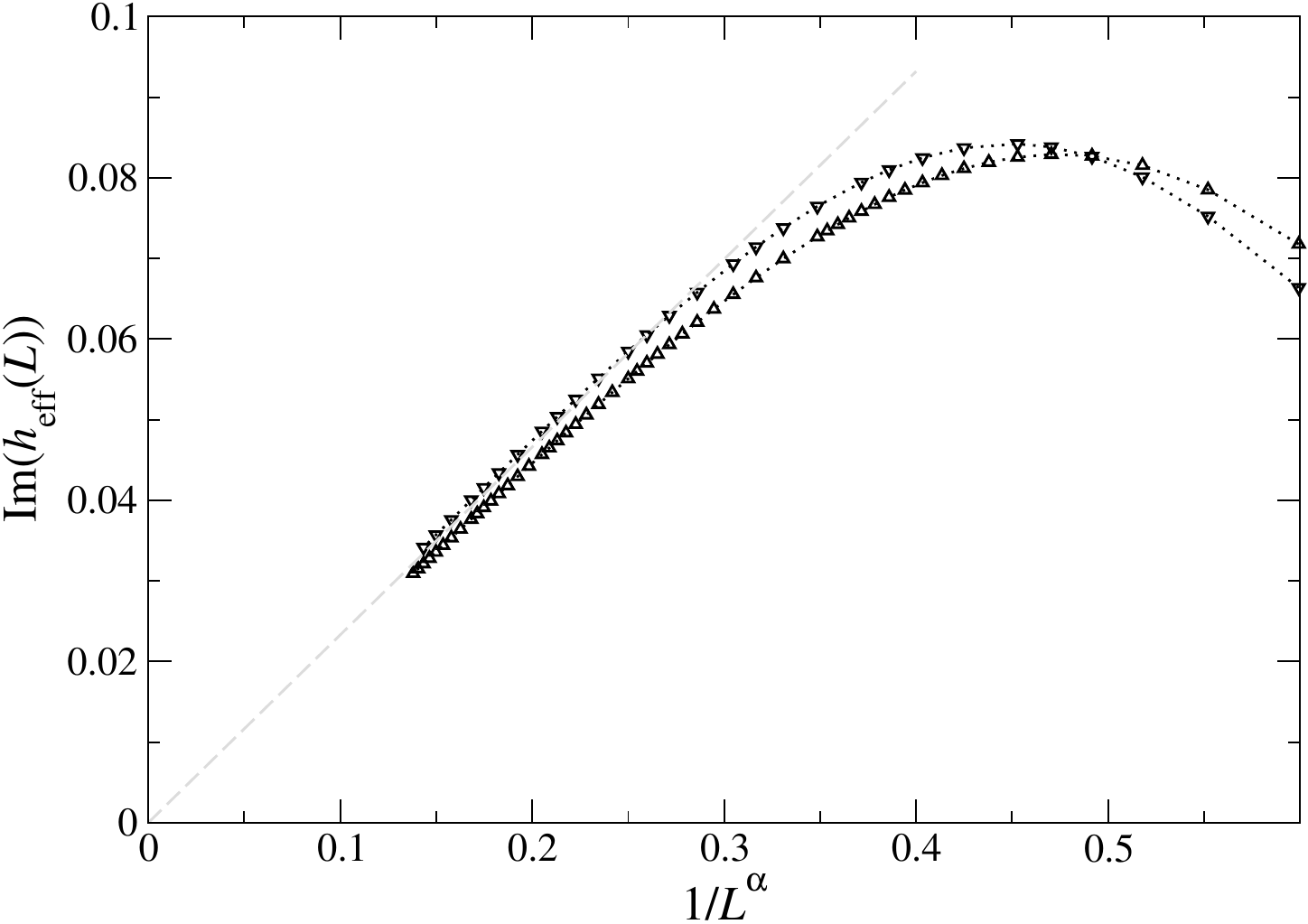}
  \hspace*{5mm}
  \end{center}
  \caption{\label{fig:X_01} Effective conformal weights
    $h_\mathrm{eff}(L)$ of the lowest states in the sector
    $(\repp;\repq)=(0;1)$: in the top panels the finite size data are
    displayed for $\gamma=2\pi/9$ (left panel) and $\gamma=2\pi/7$
    (right panel).  Red diamonds indicate the conformal weights
    $h_{(0;1)}=\gamma/\pi+k$, $k=0,1,2,3$.
    Note that two of the descendent fields correspond to complex energies
    in the finite size spectrum. For these the real part of the conformal
    weight is shown (marked by $\triangle$ and $\triangledown$).
    The lower left panel shows finite size data of the lowest states
    which extrapolate to $h_{(0;1)}=\gamma/\pi$ in the thermodynamic
    limit (indicated by red $\diamond$).  The dashed lines indicate
    the conjectured corrections to scaling $\propto L^{-\alpha}$ with
    $\alpha=\gamma/(\pi-\gamma)$. Similarly, in the lower right panel
    the vanishing of the imaginary parts of the complex energies for
    $\gamma=2\pi/9$ with the same subleading power law
    $\propto L^{-\alpha}$ is shown.}
\end{figure}

Next we have studied the lowest states in the sector
$(\repp;\repq)=(0;1)$, see Figure~\ref{fig:X_01}: in the thermodynamic
limit the numerical estimates for the conformal weights
\begin{equation}
    h_{\text{eff}}(L) = \frac{L}{\pi v_F}\,\left(E(L)-L\,\epsilon_\infty - f_\infty^{(OSp)}\right)\,,
\end{equation}
converge to the $h_{(0;1)}=\gamma/\pi$ (plus positive integers for the
descendent fields) in accordance with our proposal Eq.~(\ref{OPE}).
For all states we observe subleading corrections to scaling in the
conformal weights vanishing as
$h_{\text{eff}}(L)-h_{(0;1)}\propto L^{-\alpha}$ with an exponent
$\alpha=\gamma/(\pi-\gamma)$.  Such power law corrections have also
been observed in the periodic $\ospq$ model \cite{FrHM19}. Typically
they originate from the presence of irrelevant operators with scaling
dimensions larger than two \cite{Card86a}.  Note that this
perturbation becomes marginal ($\alpha\to0$) in the isotropic limit
where it becomes the source of the logarithmic fine structure observed
in the conformal spectrum of the $OSp(3|2)$ chain with free boundaries
\cite{FrMa22}.  We note that the finite size estimates of the
conformal weights for some of the descendent fields are complex.  We
find, however, that the imaginary parts are again subleading
corrections to scaling which vanish as $L^{-\alpha}$ in the
thermodynamic limit, see Fig.~\ref{fig:X_01}.

\begin{figure}[t]
  \begin{center}
      \includegraphics[width=0.45\textwidth]{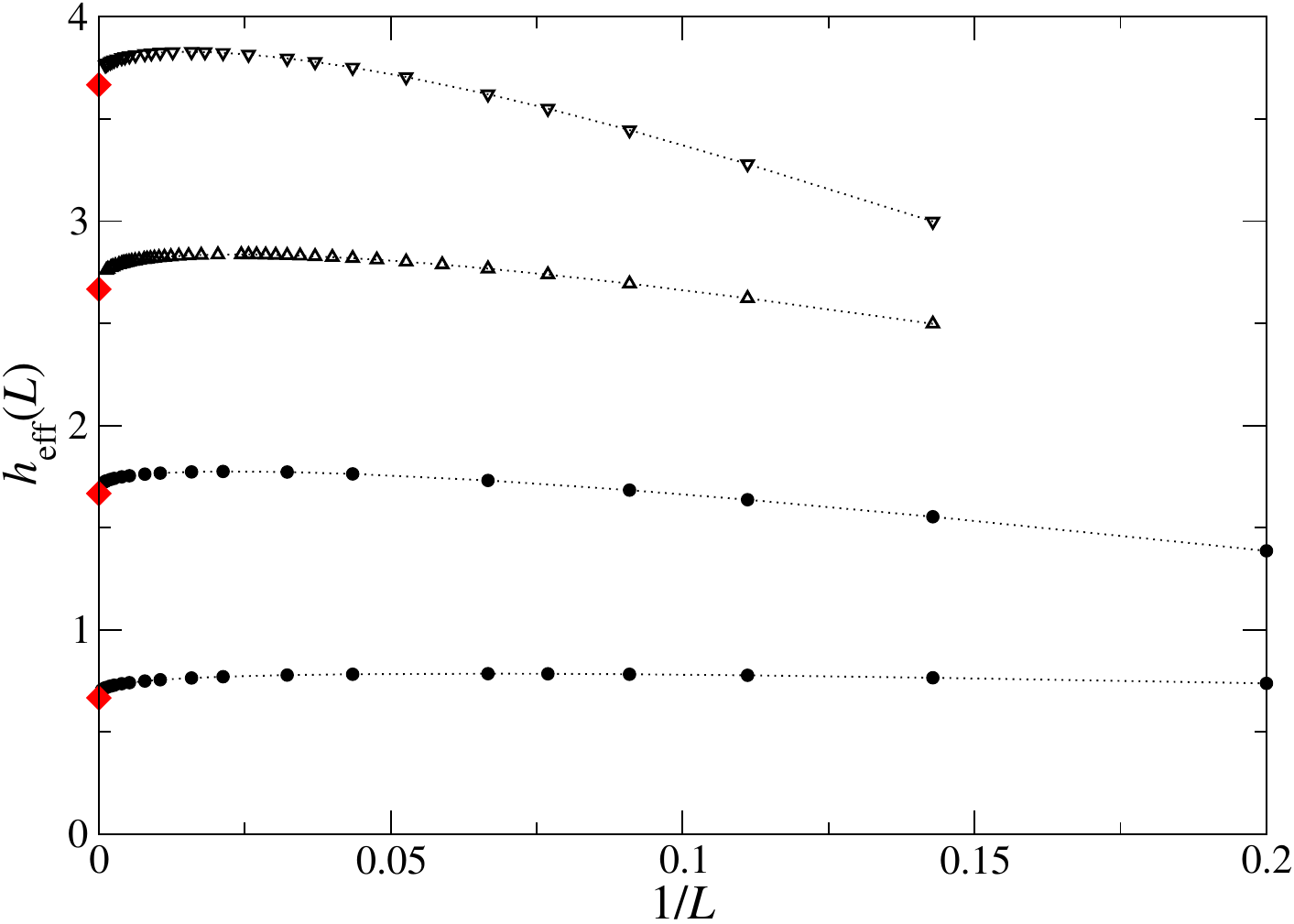}
      \hspace*{5mm}\includegraphics[width=0.45\textwidth]{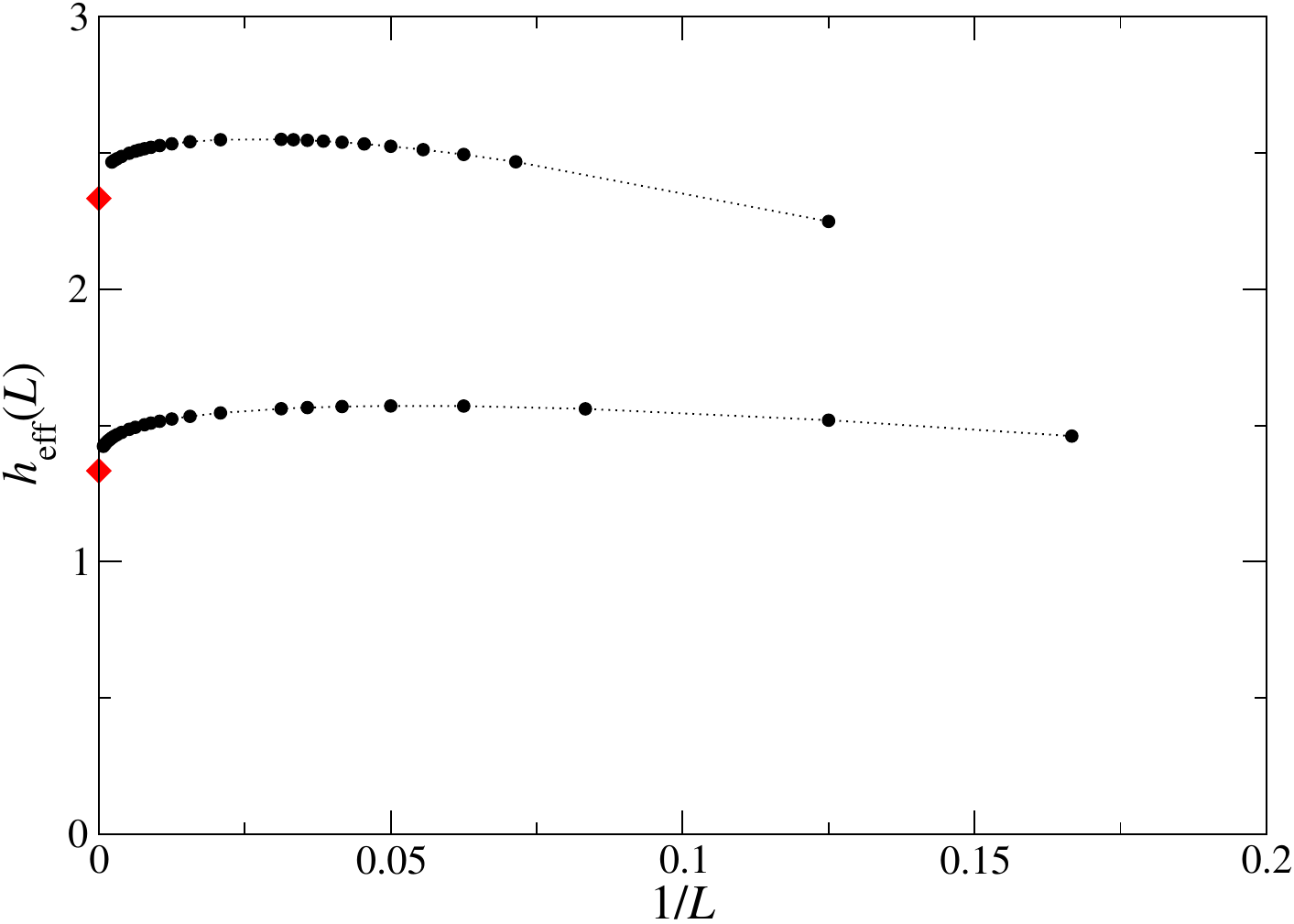}
  \end{center}
  \caption{\label{fig:X_0q}Finite size data for the primary and some
    descendents in the $(\repp;\repq)=(0;\frac32)$- (left panel) and
    $(0;2)$-sector (right panel) for $\gamma=2\pi/9$.  Red diamonds
    indicate the conformal weights
    $h_{(0;\frac32)} = 3\gamma/\pi + k$, $k=0,1,2,3$ and
    $h_{(0;2)} = 6\gamma/\pi + k$, $k=0,1$. Note that the level-2 and
    -3 descendents of the $(0;\frac32)$ appear as complex energies in
    the finite size spectrum (triangular symbols).  Similar as for the
    complex levels in the $(0;1)$-sector shown in Fig.~\ref{fig:X_01}
    the imaginary part of the effective conformal weights vanishes
    with the subleading power law in the thermodynamic limit.}
\end{figure}

Proceeding in the same way for the lowest states in the sectors
$(0;\repq)$ with $\repq=\frac32$ and $2$ we find that the finite size
data converge to the proposed values $h_{(0;\frac32)}=3\gamma/\pi$ and
$h_{(0;2)}=6\gamma/\pi$ again with subleading power law corrections
$\propto L^{-\alpha}$, $\alpha={\gamma/(\pi-\gamma)}$, see
Figure~\ref{fig:X_0q}.  Again, the energies of the level-2 and -3
descendents of the $(0;\frac32)$ primary are complex for finite chains
with their imaginary parts vanishing as $L^{-\alpha}$ the
thermodynamic limit.

Note that the conformal weights of the $(0;\repq)$-primaries for
$\repq=1,\frac32,2,\dots$ degenerate with the $(0;0)$- and
$(0;\frac12)$-vacua of the even and odd length chain with $h=0$ in the
thermodynamic limit.  The apparent degeneracy is lifted by the
subleading corrections which become logarithmic in the limit
$\gamma\to0$.  This is consistent with what has been found for the
periodic model \cite{MaNR98,JaRS03,FrMa15} and the observation in our
previous study of the $OSp(3|2)$-symmetric superspin chain subject to
free boundary conditions \cite{FrMa22} where we found
\begin{equation}
    \left.h_{(0;\repq)}\right|_{\gamma=0} \simeq \frac{2\repq(2\repq-1)}{\log L}\,.
\end{equation}

\begin{figure}[ht]
  \begin{center}
      \includegraphics[width=0.45\textwidth]{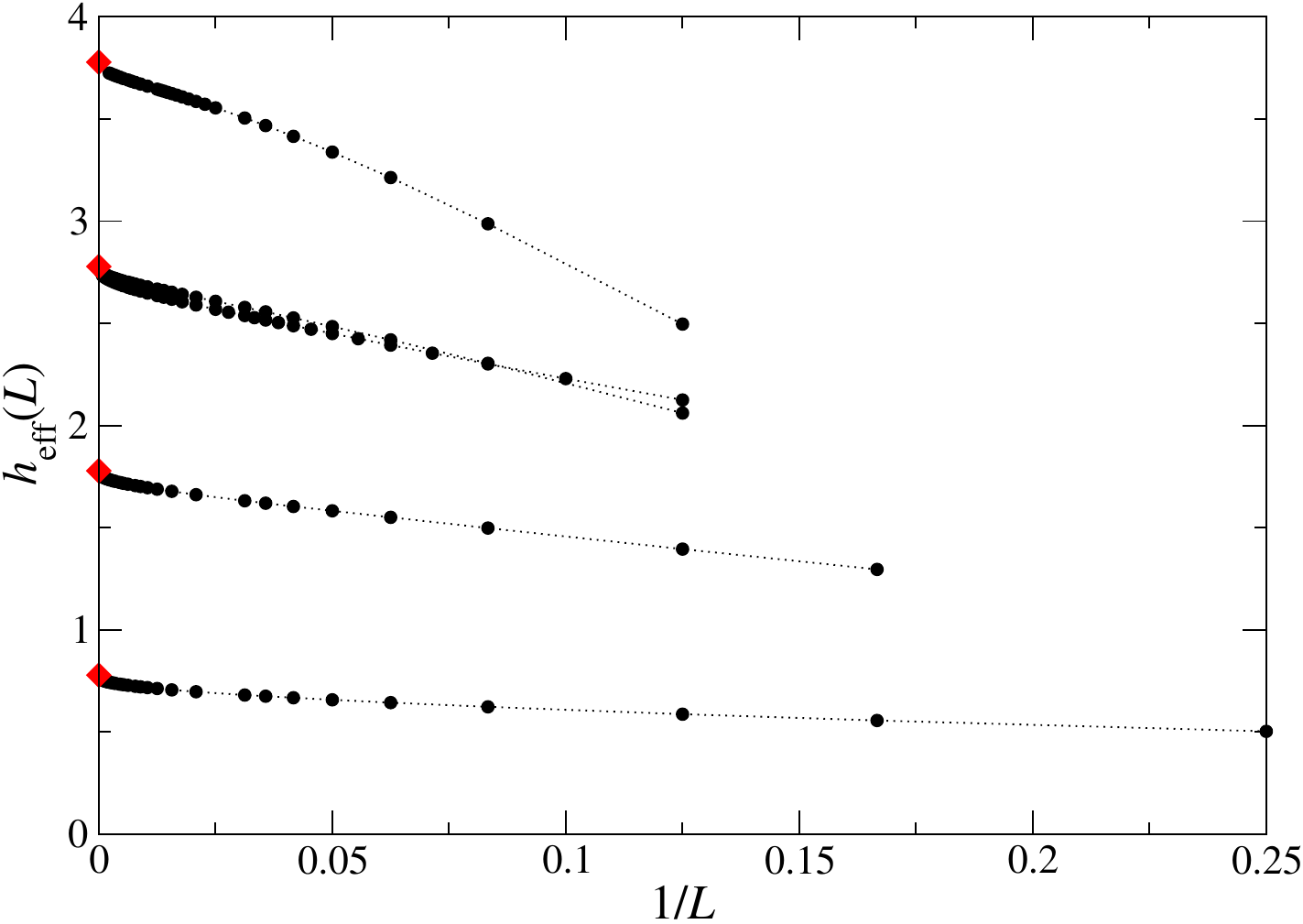}
      \hspace*{5mm}\includegraphics[width=0.45\textwidth]{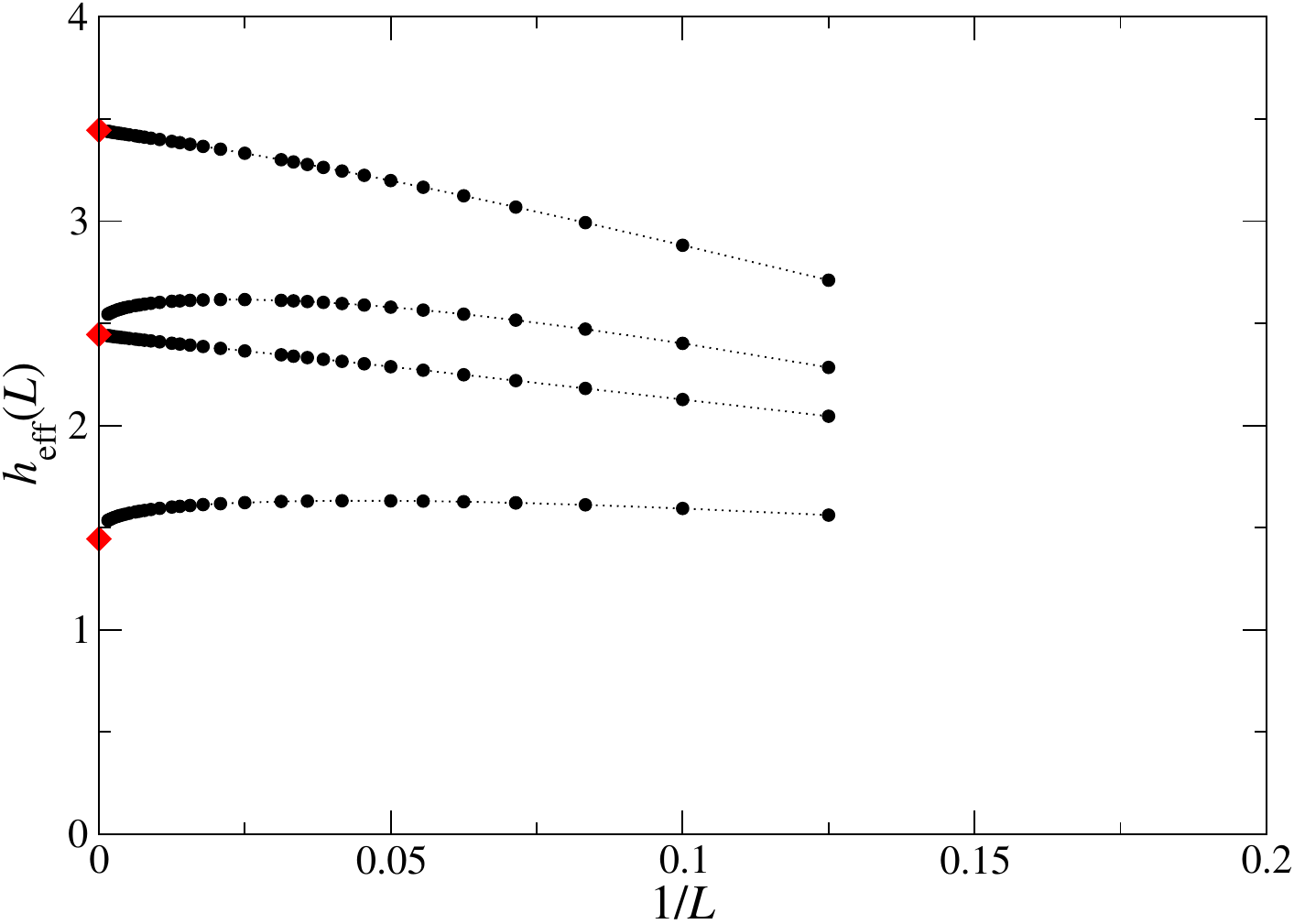}
  \end{center}
  \caption{\label{fig:X_1q}Finite size data for the primary and some
    descendents in the $(\repp;\repq)=(1;\frac12)$- (left panel) and
    $(1;\frac32)$-sector (right panel) for $\gamma=2\pi/9$.  Red
    diamonds indicate the conformal weights
    $h_{(1;\frac12)} = 1-\gamma/\pi + k$, $k=0,1,2,3$ and
    $h_{(1;\frac32)}=1+2\gamma/\pi + k$, $k=0,1,2$.}
\end{figure}

A similar behaviour can be observed for the conformal weights of the
$(1;\repq)$-primaries: for $\repq=\frac12$ and $\frac32$ these are
realized in the finite size spectrum of even $L$ chains.  The
corresponding finite size data for the primaries and some descendents
are shown in Figure~\ref{fig:X_1q}.
In the thermodynamic limit the effective conformal weights of the
primaries in these sectors converge to $h_{(1,\frac12)}=1-\gamma/\pi$
and $h_{(1,\frac32)}=1+2\gamma/\pi$, respectively, in agreement with
our proposal (\ref{OPE}).  Again one observes power law corrections to
scaling vanishing as $L^{-\alpha}$.

\begin{figure}[ht]
  \begin{center}
      \includegraphics[width=0.45\textwidth]{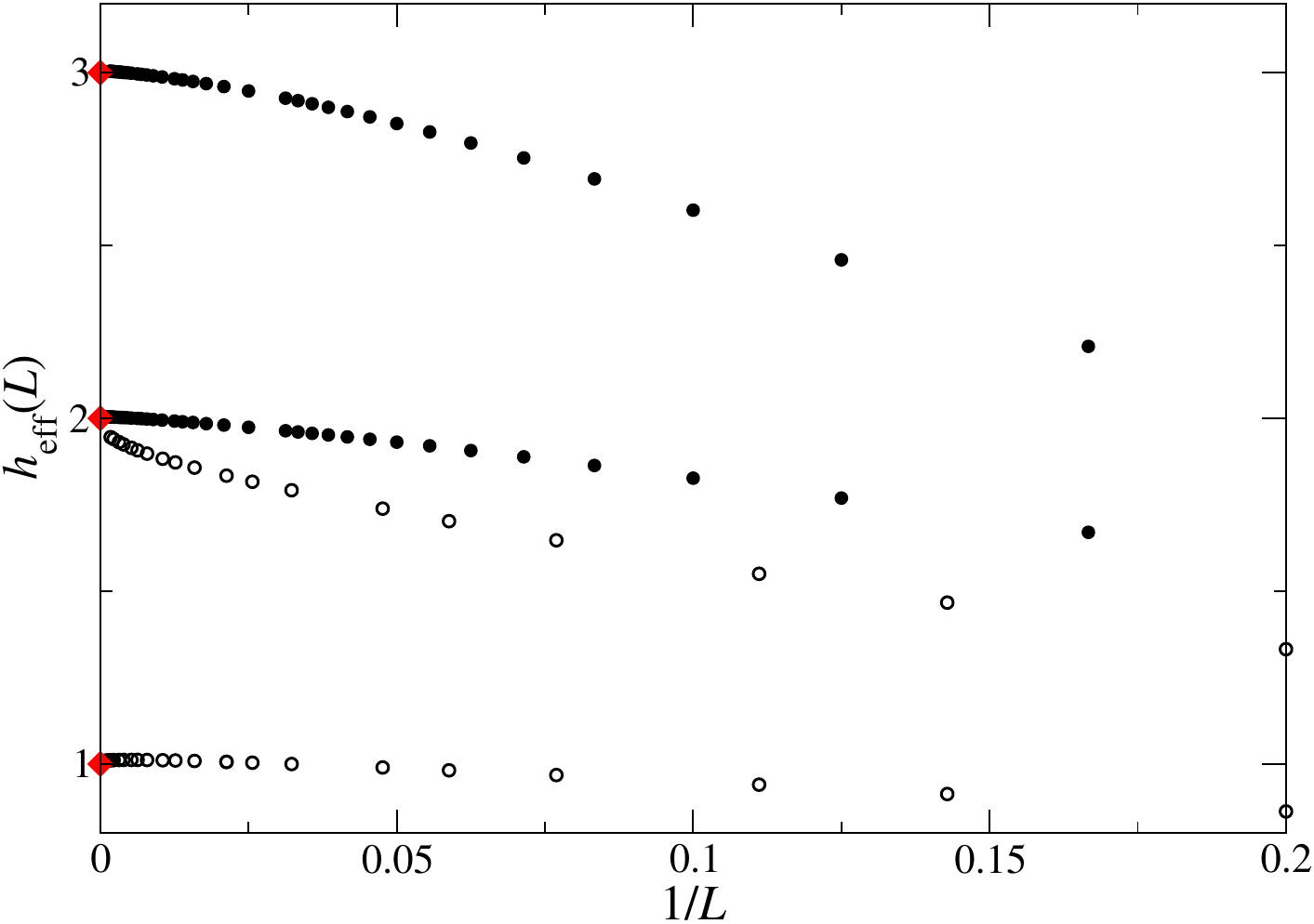} \hspace*{5mm}
      \includegraphics[width=0.45\textwidth]{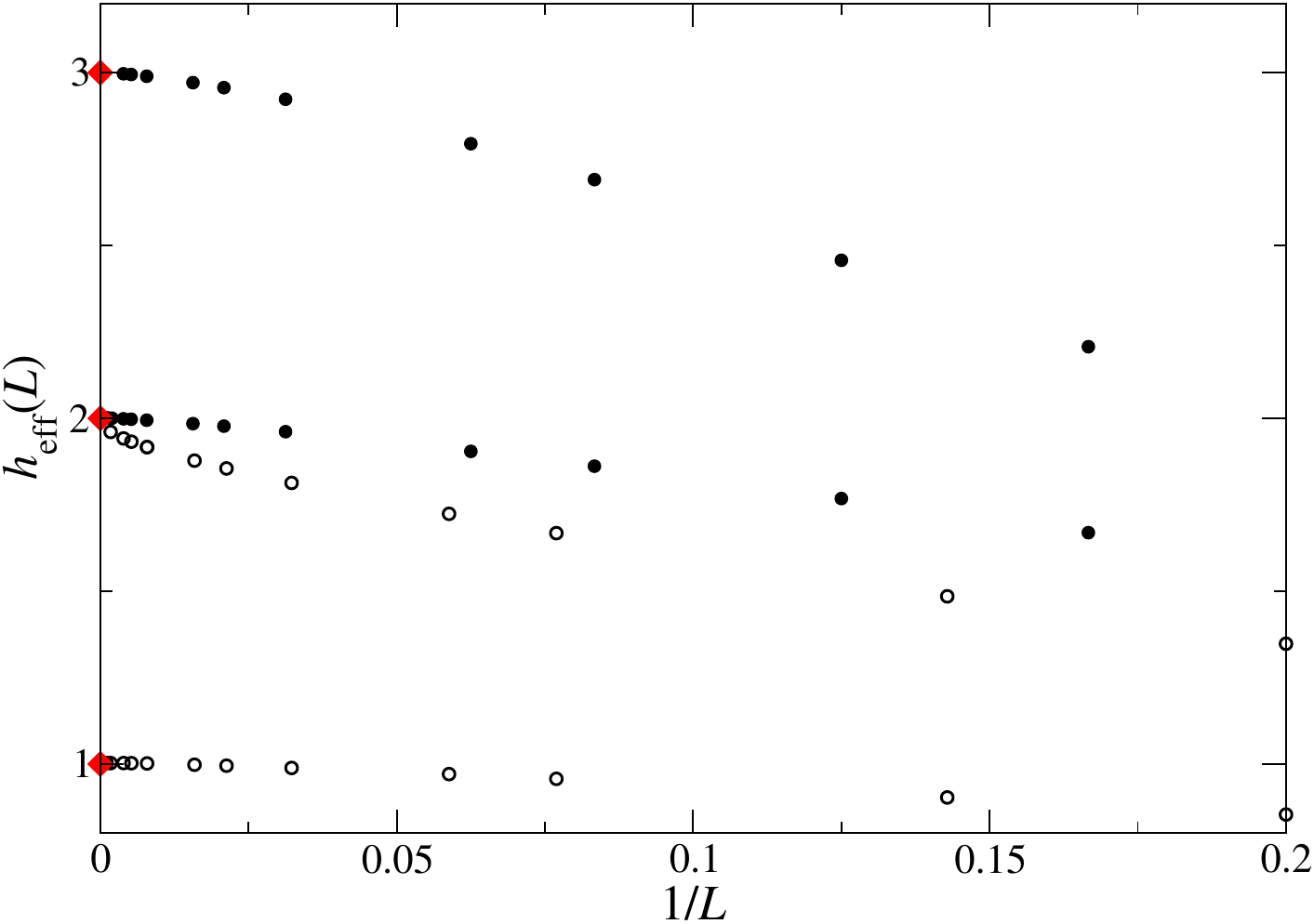}
  \end{center}
  \caption{\label{fig:X_11}Finite size data for the primary and some
    descendents in the $(\repp;\repq)=(1;1)$-sector for
    $\gamma=2\pi/9$ (left panel) and $\gamma=2\pi/7$ (right panel).
    Red diamonds indicate the conformal weights $h_{(1;1)} = 1+k$,
    $k=0,1,2$. Open (filled) symbols denote data for odd (even)
    lattice length.}
\end{figure}

As mentioned in the introduction the $(1;1)$ representation is
atypical and appears chains of length with either parity as part of
indecomposables.  According to (\ref{OPE}) the conformal weights of
operators for this representation are integers, independent of the
anisotropy $\gamma$.  This is supported by the finite size estimates
of the lowest two conformal weights for both odd and even length
presented in Fig.~\ref{fig:X_11}.
For $L$ odd the ground state in this sector extrapolates to
$h_{(1;1)}=1$, independent of the anisotropy. The lowest $(1;1)$-level
observed for even length corresponds to a field with conformal weight
$h=2$ which we may identify with the stress tensor.

As in the case of the $\repp=0$ the
$(\repp;\repq)=(1,\repq)$-primaries degenerate to give the integer
conformal weight $\left.h_{(1;\repq)}\right|_{\gamma=0}=1$ in the
thermodynamic limit.  The subleading corrections become logarithmic
consistent with the critical behaviour observed for the isotropic
$OSp(3|2)$ model \cite{FrMa15}.

\begin{figure}[ht]
  \begin{center}
      \includegraphics[width=0.45\textwidth]{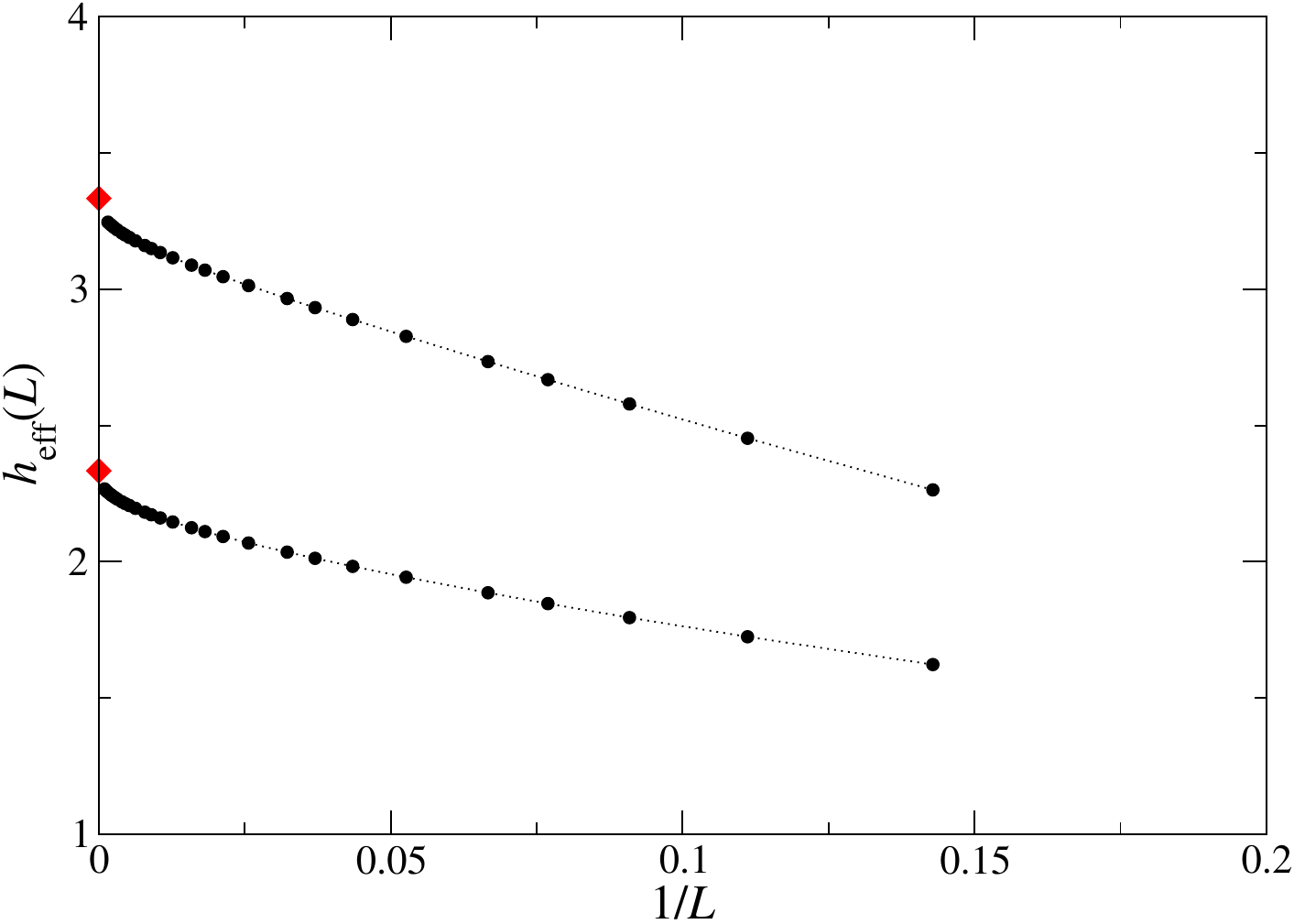}
      \hspace*{5mm}\includegraphics[width=0.45\textwidth]{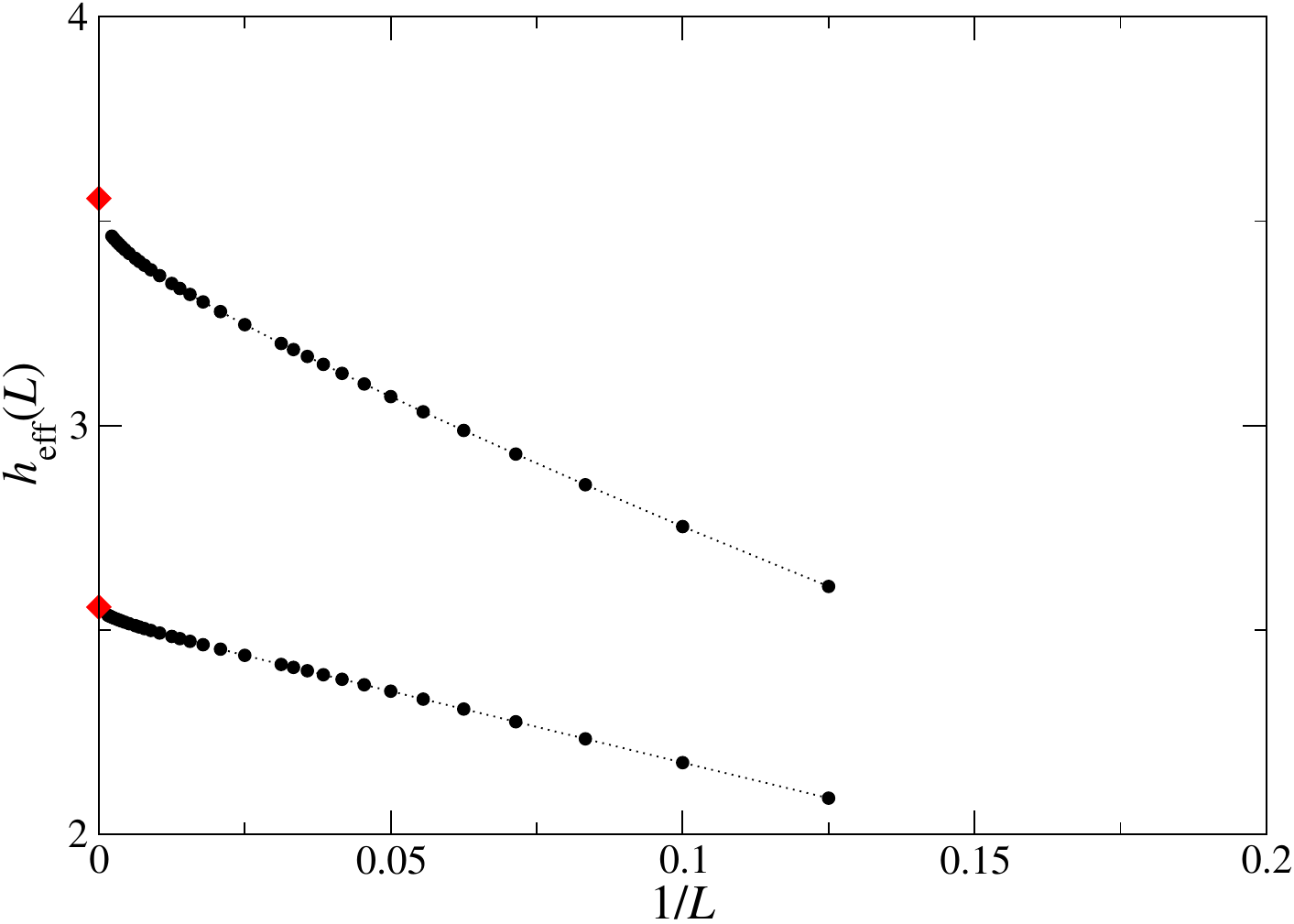}
  \end{center}
  \caption{\label{fig:X_2q}Finite size data for the primary and first
    descendents in the $(\repp;\repq)=(2;\frac12)$- (left panel) and
    $(2;1)$-sector (right panel) for $\gamma=2\pi/9$.  Red diamonds
    indicate the conformal weights
    $h_{(2;\frac12)} = 3-3\gamma/\pi + k$ and
    $h_{(2;1)}=3-2\gamma/\pi + k$, both for $k=0,1$.}
\end{figure}

Finally, we have studied the finite size spectrum of levels in the
$(\repp;\repq)=(2;\repq)$-representations, see Figure~\ref{fig:X_2q}.
For $\repq=\frac12,1$ the effective conformal weights of primaries and
first descendents extrapolate to the proposal (\ref{OPE}) for
$h_{(2;\repq)}$ and $h_{(2;\repq)}+1$.  As for the sectors with
$\repp=0,1$ they degenerate in the isotropic limit $\gamma\to0$,
consistent with what has been found previously \cite{JaRS03,FrMa15}.

\section{Summary and Conclusion}
Based on our analysis of the finite-size spectrum of the
$\ospq$-invariant superspin chain in the previous section together
with the spectral correspondence to the quantum group invariant $XXZ$
spin-$1$ spin chain we arrive at the proposal (\ref{OPE}) for the
operator content of the former.  We emphasize that this proposal is
consistent with previous results for the isotropic superspin chain:
for $\gamma\to0$ the conformal weights $h_{(p;q)}=\repp(\repp+1)/2$
coincide with those identified in the periodic model \cite{FrMa15}.
Moreover, the amplitudes of the subleading logarithmic corrections in
the $(0;\repq)$-multiplets of the isotropic model with free boundaries
\cite{FrMa22} have the same $\repq$-dependence as (\ref{OPE}).

\begin{figure}[t]
  \begin{center}
      \includegraphics[width=0.45\textwidth]{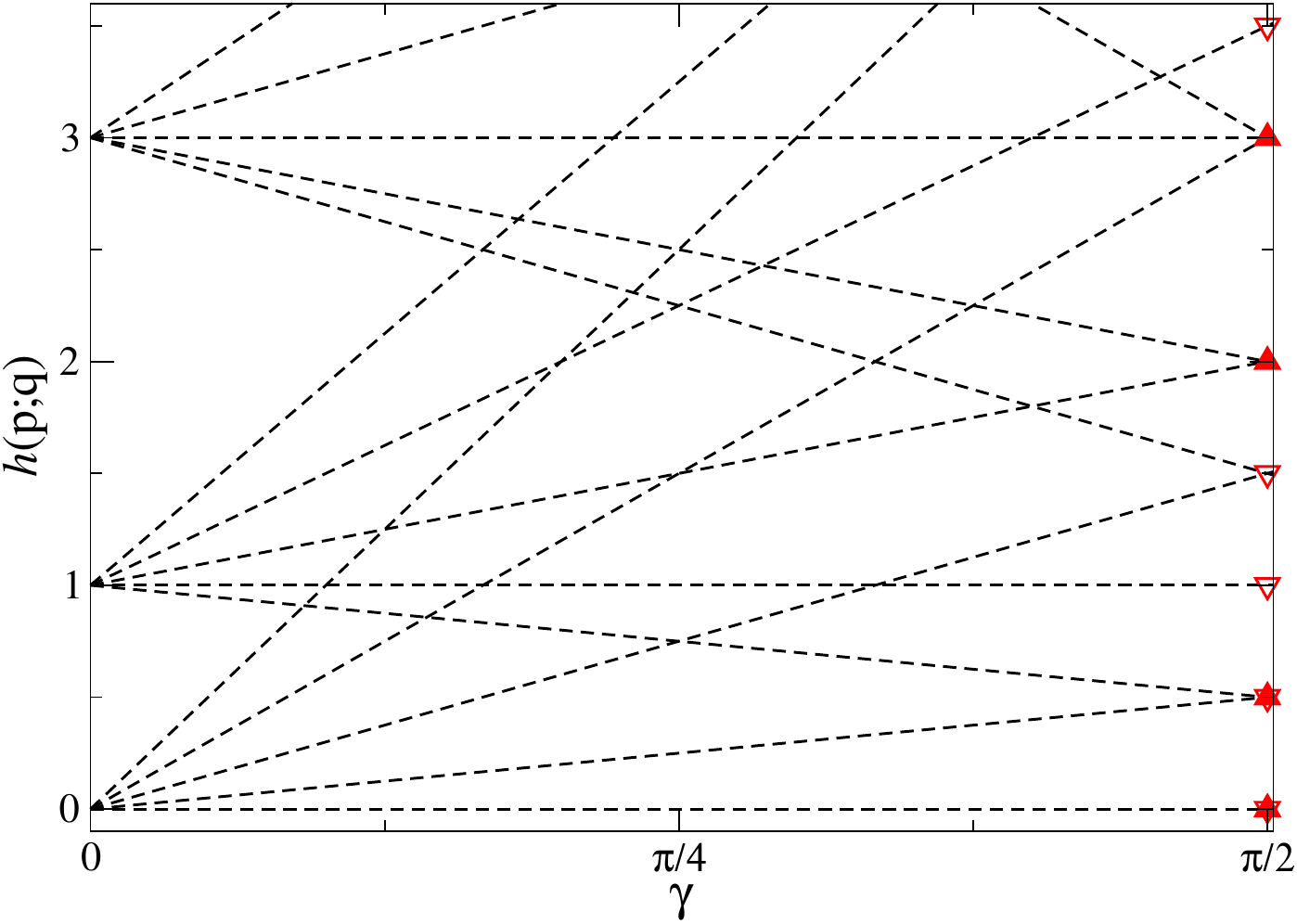}
  \end{center}
  \caption{\label{fig:CFT_flow}Conformal weights (\ref{OPE}) of
    primaries $(\repp;\repq)$ with $\repp=0,1,2,3$, $\repq\leq\frac52$
    in the $\ospq$-invariant superspin as function of the anisotropy:
    the degeneracies of weights with different $\repq$ in the
    Goldstone phase of the isotropic model are lifted for any
    $\gamma>0$.  For $\gamma=\pi/2$ the possible weights coincide with
    those of the quantum group invariant $XXZ$ spin-$1$ chain
    (\ref{eq:xxzq_KoSa}) and (\ref{eq:xxzq_Ising}) (red triangles,
    filled (open) for even (odd) chain lengths). }
\end{figure}

For the models with $\gamma>0$ a similar link to the $q$-deformed
superspin chain with periodic boundary conditions can not be
established.  In view of what is known for the periodic model the most
striking property of the quantum group invariant one studied in the
present paper is its purely discrete spectrum of conformal weights,
see Fig.~\ref{fig:CFT_flow},
while there have been indications for the presence of contiuous
components in the periodic model \cite{FrHM19}.  This resembles the
behaviour observed in the open staggered six-vertex model: depending
on the choice of boundary conditions this system flows to different
fixed points with the non-compact one being unstable, see
Refs.~\cite{RoJS19,RoJS21,FrGe22,FrGe23}.

At this point we remark that solutions of the boundary Yang-Baxter
equation for the $q$-deformed $OSp(3|2)$ spin chain beyond the quantum
group invariant one have been investigated in the literature: in
\cite{Lima09} the author claims that there exist two additional
diagonal solutions (the most amenable ones for a Bethe ansatz
solution). Just as the constant boundary matrices $K^{\pm}$ considered
here, none of these does contain any free parameter though.  It may
well be that non-compact degrees of freedom are present in the scaling
limit of the superspin chain with one of these additional boundary
conditions (or a combination of two of the three known ones).  In the
absence of a free parameter such integrable models would however be
isolated points in the space of boundary parameters and the system
sizes needed for tests of whether or not the corresponding critical
properties are robust against perturbations appear to be out of reach
for numerical approaches.

Another obstacle is the lack of knowledge on the conformal spectrum of
the periodic superspin chain where the connection to the spectrum of
the periodic Zamolodchikov-Fateev spin-$1$ chain for $\gamma=\pi/2$ is
limited to certain charge-sectors.  While this has been taken as a
hint that the fields appearing in the effective low energy theory for
the periodic $\ospq$ model may be composites of two Gaussian fields
and an Ising operator the preliminary proposal for the scaling
dimensions of the periodic superspin chain is far from a complete
description of the full operator content of the critical model.  In
particular, the dependence of the scaling dimensions on the two
possible twist angles related to the conserved $U(1)$ charges has not
been addressed.

\section*{Data availability}
The Bethe roots and finite size spectral data used in
Figures~\ref{fig:xxz1_odd}--\ref{fig:X_2q} are available from the
Research Data Repository at Leibniz Universität Hannover
\cite{FrMa23.data}.

\begin{acknowledgments}
  Funding for this work has been provided by the Deutsche
  Forschungsgemeinschaft under grant No.\ Fr~737/9-2 as part of the
  research unit \emph{Correlations in Integrable Quantum Many-Body
    Systems} (FOR2316). MJM was partially supported by the Brazilian
  agency CNPq under grant no.\ 305617/2021-4.
\end{acknowledgments}

\appendix
\section{The quantum group invariant Hamiltonian}
\label{app:HAMq}
Here we write the the quantum group invariant Hamiltonian (\ref{HAM})
in the following form,
\begin{equation}
H= \sum_{j=1}^{L-1} H^b_{j,j+1} + H^s_{L,1}
\end{equation}
where $H^b_{j,j+1}$ represents the two-body bulk term and $H^s_{L,1}$
encodes the surface interactions.

The surface term can be represented in the terms of azimuthal bosonic
operators as follows
\begin{equation}
H^s_{L,1}= -i(\tau^z_{L} -\tau^{z}_{1}) -2i(\sigma^{z}_L-\sigma^{z}_{1})
\end{equation}

The bulk term encodes both the bosonic and the fermionic generators
and it is given by,
\begin{eqnarray}
H^b_{j,j+1} &=& -\frac{\left( C_{j,j+1}-3C_{j,j+1}^2 \right)}{\sin(\gamma)}+ \delta_1 \tau^{z}_j \tau^{z}_{j+1} +\delta_2 \sigma^{z}_j \sigma^{z}_{j+1} 
+\sin(\gamma) \left(\tau^z_j \tau^z_{j+1} \right)^2 +\delta_3\left(\sigma_j^{z} \sigma_{j+1}^{z}\right)^2  \nonumber \\
	&+&\frac{\tan(\frac{\gamma}{2})}{2}\left[(\tau_j^{+} \tau_{j+1}^{-} +\tau_j^{-} \tau_{j+1}^{+})^2+2(\tau_j^z)^2+2(\tau_{j+1}^z)^2  
	+24(\sigma_j^z)^2+24(\sigma_{j+1}^z)^2 \right] \nonumber \\
&-&i\left[2+\cos(\gamma)\right] \tau_j^{z} \tau_{j+1}^{z} (\tau_{j+1}^z-\tau_j^{z}) 
+8i\cos(\gamma)\sigma_j^{z} \sigma_{j+1}^{z} (\sigma_{j+1}^z-\sigma_j^{z}) \nonumber \\ 
&-&t_1^{-} \left(c_j^{+} c_{j+1}^{-} c_j^{-} c_{j+1}^{+} +f_j^{-} f_{j+1}^{+} f_j^{+} f_{j+1}^{-} +\tau_j^{z} \tau_{j+1}^{z} c_j^{+} c_{j+1}^{-} 
-4\sigma_j^{z}\sigma_{j+1}^{z} c_j^{-} c^{+}_{j+1} \right) \nonumber \\
&-&t_1^{+} \left(c_j^{-} c_{j+1}^{+} c_j^{+} c_{j+1}^{-} +f_j^{+} f_{j+1}^{-} f_j^{-} f_{j+1}^{+} -\tau_j^{z} \tau_{j+1}^{z} c_j^{-} c_{j+1}^{+} 
+4\sigma_j^{z}\sigma_{j+1}^{z} c_j^{+} c^{-}_{j+1} \right) \nonumber \\
	&-&\frac{t_2^{+}}{2} \left[(\tau_j^{+}\tau_{j+1}^{-}+\tau_j^{-}\tau_{j+1}^{+})\tau_j^z\tau_{j+1}^{z}-(\tau_j^{+}\tau_{j+1}^{-}+\tau_j^{-}\tau_{j+1}^{+})f_j^{-} f_{j+1}^{+} -8\sigma_j^{z} \sigma_{j+1}^{z} d_j^{-} d_{j+1}^{+} \right] \nonumber \\
	&-&\frac{t_2^{-}}{2} \left[\tau_j^{z}\tau_{j+1}^{z}(\tau_j^{+}\tau_{j+1}^{-}+\tau_j^{-}\tau_{j+1}^{+})+(\tau_j^{+}\tau_{j+1}^{-}+\tau_j^{-}\tau_{j+1}^{+})f_j^{+} f_{j+1}^{-} +8\sigma_j^{z} \sigma_{j+1}^{z} d_j^{+} d_{j+1}^{-} \right] \nonumber \\
&+&t_3^{+}\left(\tau_j^{z}\tau_{j+1}^{z} f_j^{-} f_{j+1}^{+}-4 \sigma_j^{z}\sigma_{j+1}^{z} f_j^{+} f_{j+1}^{-} \right) 
-t_3^{-}\left(\tau_j^{z}\tau_{j+1}^{z} f_j^{+} f_{j+1}^{-}-4 \sigma_j^{z}\sigma_{j+1}^{z} f_j^{-} f_{j+1}^{+} \right) \nonumber \\
	&+& 4i \cos(\frac{\gamma}{2}) \left(f_j^{-} f_{j+1}^{+} d_j^{-} d_{j+1}^{+} -d_j^{-}d_{j+1}^{+} f_j^{-}f_{j+1}^{+} \right) -2\left(\tan(\frac{\gamma}{2}) 
+\frac{2}{\sin(\gamma)} \right) \mathrm{I_j} \mathrm{I_{j+1}}
\end{eqnarray}
where $I_j$ denotes the $5 \times 5$ identity acting on the $jth$
lattice site. The dependence of the coupling parameters on the
anisotropy $\gamma$ is,
\begin{eqnarray}
	&&\delta_1= 2 \sin^2(\frac{\gamma}{2}){\tan(\frac{\gamma}{2})},~~\delta_2=4\left[2+\cos(\gamma)\right] \tan(\frac{\gamma}{2}),~~\delta_3=-16\left[6+\cos(\gamma)\right] \tan(\frac{\gamma}{2}) \nonumber \\
&& t_1^{\pm}= \pm i \frac{\exp(\pm i\frac{\gamma}{2})}{\cos(\frac{\gamma}{2})},~~
t_2^{\pm}= \pm i\frac{\exp(\pm i\frac{\gamma}{2})}{\cos(\frac{\gamma}{2})}\left(1+\exp(\pm i\frac{\gamma}{4})\frac{\cos(\frac{\gamma}{2})}{\cos(\frac{\gamma}{4})}\right) \nonumber \\
&& t_3^{\pm}= \pm i \frac{\exp(\pm i\frac{\gamma}{2})}{\cos(\frac{\gamma}{2})}\left(1+2\exp(\pm i\frac{\gamma}{2})\cos(\frac{\gamma}{2}) \right)
\end{eqnarray}
We finally remark that one may use the commutation relations among the
bosonic and fermionic generators of the $OSp(3|2)$ superalgebra to
write alternative expressions to the bulk Hamiltonian.
\bibliographystyle{apsrev4-1}

\begin{thebibliography}{52}%
\makeatletter
\providecommand \@ifxundefined [1]{%
 \@ifx{#1\undefined}
}%
\providecommand \@ifnum [1]{%
 \ifnum #1\expandafter \@firstoftwo
 \else \expandafter \@secondoftwo
 \fi
}%
\providecommand \@ifx [1]{%
 \ifx #1\expandafter \@firstoftwo
 \else \expandafter \@secondoftwo
 \fi
}%
\providecommand \natexlab [1]{#1}%
\providecommand \enquote  [1]{``#1''}%
\providecommand \bibnamefont  [1]{#1}%
\providecommand \bibfnamefont [1]{#1}%
\providecommand \citenamefont [1]{#1}%
\providecommand \href@noop [0]{\@secondoftwo}%
\providecommand \href [0]{\begingroup \@sanitize@url \@href}%
\providecommand \@href[1]{\@@startlink{#1}\@@href}%
\providecommand \@@href[1]{\endgroup#1\@@endlink}%
\providecommand \@sanitize@url [0]{\catcode `\\12\catcode `\$12\catcode
  `\&12\catcode `\#12\catcode `\^12\catcode `\_12\catcode `\%12\relax}%
\providecommand \@@startlink[1]{}%
\providecommand \@@endlink[0]{}%
\providecommand \url  [0]{\begingroup\@sanitize@url \@url }%
\providecommand \@url [1]{\endgroup\@href {#1}{\urlprefix }}%
\providecommand \urlprefix  [0]{URL }%
\providecommand \Eprint [0]{\href }%
\providecommand \doibase [0]{http://dx.doi.org/}%
\providecommand \selectlanguage [0]{\@gobble}%
\providecommand \bibinfo  [0]{\@secondoftwo}%
\providecommand \bibfield  [0]{\@secondoftwo}%
\providecommand \translation [1]{[#1]}%
\providecommand \BibitemOpen [0]{}%
\providecommand \bibitemStop [0]{}%
\providecommand \bibitemNoStop [0]{.\EOS\space}%
\providecommand \EOS [0]{\spacefactor3000\relax}%
\providecommand \BibitemShut  [1]{\csname bibitem#1\endcsname}%
\let\auto@bib@innerbib\@empty
\bibitem [{\citenamefont {Essler}\ \emph {et~al.}(2005)\citenamefont {Essler},
  \citenamefont {Frahm},\ and\ \citenamefont {Saleur}}]{EsFS05}%
  \BibitemOpen
  \bibfield  {author} {\bibinfo {author} {\bibfnamefont {F.~H.~L.}\
  \bibnamefont {Essler}}, \bibinfo {author} {\bibfnamefont {H.}~\bibnamefont
  {Frahm}}, \ and\ \bibinfo {author} {\bibfnamefont {H.}~\bibnamefont
  {Saleur}},\ }\href {\doibase 10.1016/j.nuclphysb.2005.01.021} {\bibfield
  {journal} {\bibinfo  {journal} {Nucl. Phys. B}\ }\textbf {\bibinfo {volume}
  {712 [FS]}},\ \bibinfo {pages} {513} (\bibinfo {year} {2005})},\ \Eprint
  {http://arxiv.org/abs/cond-mat/0501197} {cond-mat/0501197} \BibitemShut
  {NoStop}%
\bibitem [{\citenamefont {Ikhlef}\ \emph {et~al.}(2008)\citenamefont {Ikhlef},
  \citenamefont {Jacobsen},\ and\ \citenamefont {Saleur}}]{IkJS08}%
  \BibitemOpen
  \bibfield  {author} {\bibinfo {author} {\bibfnamefont {Y.}~\bibnamefont
  {Ikhlef}}, \bibinfo {author} {\bibfnamefont {J.~L.}\ \bibnamefont
  {Jacobsen}}, \ and\ \bibinfo {author} {\bibfnamefont {H.}~\bibnamefont
  {Saleur}},\ }\href@noop {} {\bibfield  {journal} {\bibinfo  {journal} {Nucl.
  Phys. B}\ }\textbf {\bibinfo {volume} {789}},\ \bibinfo {pages} {483}
  (\bibinfo {year} {2008})},\ \Eprint {http://arxiv.org/abs/cond-mat/0612037}
  {cond-mat/0612037} \BibitemShut {NoStop}%
\bibitem [{\citenamefont {Frahm}\ and\ \citenamefont {Martins}(2011)}]{FrMa11}%
  \BibitemOpen
  \bibfield  {author} {\bibinfo {author} {\bibfnamefont {H.}~\bibnamefont
  {Frahm}}\ and\ \bibinfo {author} {\bibfnamefont {M.~J.}\ \bibnamefont
  {Martins}},\ }\href {\doibase 10.1016/j.nuclphysb.2011.01.026} {\bibfield
  {journal} {\bibinfo  {journal} {Nucl. Phys. B}\ }\textbf {\bibinfo {volume}
  {847}},\ \bibinfo {pages} {220} (\bibinfo {year} {2011})},\ \Eprint
  {http://arxiv.org/abs/1012.1753} {arXiv:1012.1753} \BibitemShut {NoStop}%
\bibitem [{\citenamefont {Frahm}\ and\ \citenamefont {Martins}(2012)}]{FrMa12}%
  \BibitemOpen
  \bibfield  {author} {\bibinfo {author} {\bibfnamefont {H.}~\bibnamefont
  {Frahm}}\ and\ \bibinfo {author} {\bibfnamefont {M.~J.}\ \bibnamefont
  {Martins}},\ }\href@noop {} {\bibfield  {journal} {\bibinfo  {journal} {Nucl.
  Phys. B}\ }\textbf {\bibinfo {volume} {862}},\ \bibinfo {pages} {504}
  (\bibinfo {year} {2012})},\ \Eprint {http://arxiv.org/abs/1202.4676}
  {arXiv:1202.4676} \BibitemShut {NoStop}%
\bibitem [{\citenamefont {Vernier}\ \emph {et~al.}(2014)\citenamefont
  {Vernier}, \citenamefont {Jacobsen},\ and\ \citenamefont {Saleur}}]{VeJS14}%
  \BibitemOpen
  \bibfield  {author} {\bibinfo {author} {\bibfnamefont {{\'E}.}~\bibnamefont
  {Vernier}}, \bibinfo {author} {\bibfnamefont {J.~L.}\ \bibnamefont
  {Jacobsen}}, \ and\ \bibinfo {author} {\bibfnamefont {H.}~\bibnamefont
  {Saleur}},\ }\href@noop {} {\bibfield  {journal} {\bibinfo  {journal} {J.
  Phys. A: Math. Theor.}\ }\textbf {\bibinfo {volume} {47}},\ \bibinfo {pages}
  {285202} (\bibinfo {year} {2014})},\ \Eprint {http://arxiv.org/abs/1404.4497}
  {arXiv:1404.4497} \BibitemShut {NoStop}%
\bibitem [{\citenamefont {Vernier}\ \emph {et~al.}(2016)\citenamefont
  {Vernier}, \citenamefont {Jacobsen},\ and\ \citenamefont {Saleur}}]{VeJS16a}%
  \BibitemOpen
  \bibfield  {author} {\bibinfo {author} {\bibfnamefont {E.}~\bibnamefont
  {Vernier}}, \bibinfo {author} {\bibfnamefont {J.~L.}\ \bibnamefont
  {Jacobsen}}, \ and\ \bibinfo {author} {\bibfnamefont {H.}~\bibnamefont
  {Saleur}},\ }\href@noop {} {\bibfield  {journal} {\bibinfo  {journal} {Nucl.
  Phys. B}\ }\textbf {\bibinfo {volume} {911}},\ \bibinfo {pages} {52}
  (\bibinfo {year} {2016})},\ \Eprint {http://arxiv.org/abs/1601.01559}
  {arXiv:1601.01559} \BibitemShut {NoStop}%
\bibitem [{\citenamefont {Frahm}\ \emph {et~al.}(2019)\citenamefont {Frahm},
  \citenamefont {Hobu{\ss}},\ and\ \citenamefont {Martins}}]{FrHM19}%
  \BibitemOpen
  \bibfield  {author} {\bibinfo {author} {\bibfnamefont {H.}~\bibnamefont
  {Frahm}}, \bibinfo {author} {\bibfnamefont {K.}~\bibnamefont {Hobu{\ss}}}, \
  and\ \bibinfo {author} {\bibfnamefont {M.~J.}\ \bibnamefont {Martins}},\
  }\href {\doibase 10.1016/j.nuclphysb.2019.114697} {\bibfield  {journal}
  {\bibinfo  {journal} {Nucl. Phys. B}\ }\textbf {\bibinfo {volume} {946}},\
  \bibinfo {pages} {114697} (\bibinfo {year} {2019})},\ \Eprint
  {http://arxiv.org/abs/1906.00655} {arXiv:1906.00655} \BibitemShut {NoStop}%
\bibitem [{\citenamefont {Ikhlef}\ \emph {et~al.}(2012)\citenamefont {Ikhlef},
  \citenamefont {Jacobsen},\ and\ \citenamefont {Saleur}}]{IkJS12}%
  \BibitemOpen
  \bibfield  {author} {\bibinfo {author} {\bibfnamefont {Y.}~\bibnamefont
  {Ikhlef}}, \bibinfo {author} {\bibfnamefont {J.~L.}\ \bibnamefont
  {Jacobsen}}, \ and\ \bibinfo {author} {\bibfnamefont {H.}~\bibnamefont
  {Saleur}},\ }\href@noop {} {\bibfield  {journal} {\bibinfo  {journal} {Phys.
  Rev. Lett.}\ }\textbf {\bibinfo {volume} {108}},\ \bibinfo {pages} {081601}
  (\bibinfo {year} {2012})},\ \Eprint {http://arxiv.org/abs/1109.1119}
  {arXiv:1109.1119} \BibitemShut {NoStop}%
\bibitem [{\citenamefont {Candu}\ and\ \citenamefont {Ikhlef}(2013)}]{CaIk13}%
  \BibitemOpen
  \bibfield  {author} {\bibinfo {author} {\bibfnamefont {C.}~\bibnamefont
  {Candu}}\ and\ \bibinfo {author} {\bibfnamefont {Y.}~\bibnamefont {Ikhlef}},\
  }\href@noop {} {\bibfield  {journal} {\bibinfo  {journal} {J. Phys. A: Math.
  Theor.}\ }\textbf {\bibinfo {volume} {46}},\ \bibinfo {pages} {415401}
  (\bibinfo {year} {2013})},\ \Eprint {http://arxiv.org/abs/1306.2646}
  {arXiv:1306.2646} \BibitemShut {NoStop}%
\bibitem [{\citenamefont {Frahm}\ and\ \citenamefont {Seel}(2014)}]{FrSe14}%
  \BibitemOpen
  \bibfield  {author} {\bibinfo {author} {\bibfnamefont {H.}~\bibnamefont
  {Frahm}}\ and\ \bibinfo {author} {\bibfnamefont {A.}~\bibnamefont {Seel}},\
  }\href {\doibase 10.1016/j.nuclphysb.2013.12.015} {\bibfield  {journal}
  {\bibinfo  {journal} {Nucl. Phys. B}\ }\textbf {\bibinfo {volume} {879}},\
  \bibinfo {pages} {382} (\bibinfo {year} {2014})},\ \Eprint
  {http://arxiv.org/abs/1311.6911} {arXiv:1311.6911} \BibitemShut {NoStop}%
\bibitem [{\citenamefont {Bazhanov}\ \emph {et~al.}(2019)\citenamefont
  {Bazhanov}, \citenamefont {Kotousov}, \citenamefont {Koval},\ and\
  \citenamefont {Lukyanov}}]{BKKL19}%
  \BibitemOpen
  \bibfield  {author} {\bibinfo {author} {\bibfnamefont {V.~V.}\ \bibnamefont
  {Bazhanov}}, \bibinfo {author} {\bibfnamefont {G.~A.}\ \bibnamefont
  {Kotousov}}, \bibinfo {author} {\bibfnamefont {S.~M.}\ \bibnamefont {Koval}},
  \ and\ \bibinfo {author} {\bibfnamefont {S.~L.}\ \bibnamefont {Lukyanov}},\
  }\href {\doibase 10.1007/JHEP08(2019)087} {\bibfield  {journal} {\bibinfo
  {journal} {J. High. Energ. Phys.}\ }\textbf {\bibinfo {volume} {2019}},\
  \bibinfo {pages} {08, 087} (\bibinfo {year} {2019})},\ \Eprint
  {http://arxiv.org/abs/1903.05033} {arXiv:1903.05033} \BibitemShut {NoStop}%
\bibitem [{\citenamefont {Bazhanov}\ \emph
  {et~al.}(2021{\natexlab{a}})\citenamefont {Bazhanov}, \citenamefont
  {Kotousov},\ and\ \citenamefont {Lukyanov}}]{BaKL21}%
  \BibitemOpen
  \bibfield  {author} {\bibinfo {author} {\bibfnamefont {V.~V.}\ \bibnamefont
  {Bazhanov}}, \bibinfo {author} {\bibfnamefont {G.~A.}\ \bibnamefont
  {Kotousov}}, \ and\ \bibinfo {author} {\bibfnamefont {S.~L.}\ \bibnamefont
  {Lukyanov}},\ }\href {\doibase 10.1007/JHEP03(2021)169} {\bibfield  {journal}
  {\bibinfo  {journal} {J. High. Energ. Phys.}\ }\textbf {\bibinfo {volume}
  {2021}},\ \bibinfo {pages} {03, 169} (\bibinfo {year}
  {2021}{\natexlab{a}})},\ \Eprint {http://arxiv.org/abs/2010.10603}
  {arXiv:2010.10603} \BibitemShut {NoStop}%
\bibitem [{\citenamefont {Bazhanov}\ \emph
  {et~al.}(2021{\natexlab{b}})\citenamefont {Bazhanov}, \citenamefont
  {Kotousov}, \citenamefont {Koval},\ and\ \citenamefont {Lukyanov}}]{BKKL21a}%
  \BibitemOpen
  \bibfield  {author} {\bibinfo {author} {\bibfnamefont {V.~V.}\ \bibnamefont
  {Bazhanov}}, \bibinfo {author} {\bibfnamefont {G.~A.}\ \bibnamefont
  {Kotousov}}, \bibinfo {author} {\bibfnamefont {S.~M.}\ \bibnamefont {Koval}},
  \ and\ \bibinfo {author} {\bibfnamefont {S.~L.}\ \bibnamefont {Lukyanov}},\
  }\href {\doibase 10.1016/j.nuclphysb.2021.115337} {\bibfield  {journal}
  {\bibinfo  {journal} {Nucl. Phys. B}\ }\textbf {\bibinfo {volume} {965}},\
  \bibinfo {pages} {115337} (\bibinfo {year} {2021}{\natexlab{b}})},\ \Eprint
  {http://arxiv.org/abs/2010.10613} {arXiv:2010.10613} \BibitemShut {NoStop}%
\bibitem [{\citenamefont {Robertson}\ \emph {et~al.}(2019)\citenamefont
  {Robertson}, \citenamefont {Jacobsen},\ and\ \citenamefont
  {Saleur}}]{RoJS19}%
  \BibitemOpen
  \bibfield  {author} {\bibinfo {author} {\bibfnamefont {N.~F.}\ \bibnamefont
  {Robertson}}, \bibinfo {author} {\bibfnamefont {J.~L.}\ \bibnamefont
  {Jacobsen}}, \ and\ \bibinfo {author} {\bibfnamefont {H.}~\bibnamefont
  {Saleur}},\ }\href {\doibase 10.1007/JHEP10(2019)254} {\bibfield  {journal}
  {\bibinfo  {journal} {J. High. Energ. Phys.}\ ,\ \bibinfo {pages} {10, 254}}
  (\bibinfo {year} {2019})},\ \Eprint {http://arxiv.org/abs/1906.07565}
  {arXiv:1906.07565} \BibitemShut {NoStop}%
\bibitem [{\citenamefont {Robertson}\ \emph {et~al.}(2021)\citenamefont
  {Robertson}, \citenamefont {Jacobsen},\ and\ \citenamefont
  {Saleur}}]{RoJS21}%
  \BibitemOpen
  \bibfield  {author} {\bibinfo {author} {\bibfnamefont {N.~F.}\ \bibnamefont
  {Robertson}}, \bibinfo {author} {\bibfnamefont {J.~L.}\ \bibnamefont
  {Jacobsen}}, \ and\ \bibinfo {author} {\bibfnamefont {H.}~\bibnamefont
  {Saleur}},\ }\href {\doibase 10.1007/JHEP02(2021)180} {\bibfield  {journal}
  {\bibinfo  {journal} {J. High. Energ. Phys.}\ ,\ \bibinfo {pages} {(02) 180}}
  (\bibinfo {year} {2021})},\ \Eprint {http://arxiv.org/abs/2012.07757}
  {arXiv:2012.07757} \BibitemShut {NoStop}%
\bibitem [{\citenamefont {Frahm}\ and\ \citenamefont
  {Gehrmann}(2022)}]{FrGe22}%
  \BibitemOpen
  \bibfield  {author} {\bibinfo {author} {\bibfnamefont {H.}~\bibnamefont
  {Frahm}}\ and\ \bibinfo {author} {\bibfnamefont {S.}~\bibnamefont
  {Gehrmann}},\ }\href {\doibase 10.1007/JHEP01(2022)070} {\bibfield  {journal}
  {\bibinfo  {journal} {J. High. Energ. Phys.}\ ,\ \bibinfo {pages} {01, 070}}
  (\bibinfo {year} {2022})},\ \Eprint {http://arxiv.org/abs/2111.00850}
  {arXiv:2111.00850} \BibitemShut {NoStop}%
\bibitem [{\citenamefont {Frahm}\ and\ \citenamefont
  {Gehrmann}(2023)}]{FrGe23}%
  \BibitemOpen
  \bibfield  {author} {\bibinfo {author} {\bibfnamefont {H.}~\bibnamefont
  {Frahm}}\ and\ \bibinfo {author} {\bibfnamefont {S.}~\bibnamefont
  {Gehrmann}},\ }\href {\doibase 10.1088/1751-8121/acb29f} {\bibfield
  {journal} {\bibinfo  {journal} {J. Phys. A: Math. Theor.}\ }\textbf {\bibinfo
  {volume} {56}},\ \bibinfo {pages} {025001} (\bibinfo {year} {2023})},\
  \Eprint {http://arxiv.org/abs/2209.06182} {2209.06182} \BibitemShut {NoStop}%
\bibitem [{\citenamefont {Martins}\ \emph {et~al.}(1998)\citenamefont
  {Martins}, \citenamefont {Nienhuis},\ and\ \citenamefont {Rietman}}]{MaNR98}%
  \BibitemOpen
  \bibfield  {author} {\bibinfo {author} {\bibfnamefont {M.~J.}\ \bibnamefont
  {Martins}}, \bibinfo {author} {\bibfnamefont {B.}~\bibnamefont {Nienhuis}}, \
  and\ \bibinfo {author} {\bibfnamefont {R.}~\bibnamefont {Rietman}},\ }\href
  {\doibase 10.1103/PhysRevLett.81.504} {\bibfield  {journal} {\bibinfo
  {journal} {Phys. Rev. Lett.}\ }\textbf {\bibinfo {volume} {81}},\ \bibinfo
  {pages} {504} (\bibinfo {year} {1998})},\ \Eprint
  {http://arxiv.org/abs/cond-mat/9709051} {cond-mat/9709051} \BibitemShut
  {NoStop}%
\bibitem [{\citenamefont {Jacobsen}\ \emph {et~al.}(2003)\citenamefont
  {Jacobsen}, \citenamefont {Read},\ and\ \citenamefont {Saleur}}]{JaRS03}%
  \BibitemOpen
  \bibfield  {author} {\bibinfo {author} {\bibfnamefont {J.~L.}\ \bibnamefont
  {Jacobsen}}, \bibinfo {author} {\bibfnamefont {N.}~\bibnamefont {Read}}, \
  and\ \bibinfo {author} {\bibfnamefont {H.}~\bibnamefont {Saleur}},\
  }\href@noop {} {\bibfield  {journal} {\bibinfo  {journal} {Phys. Rev. Lett.}\
  }\textbf {\bibinfo {volume} {90}},\ \bibinfo {pages} {090601} (\bibinfo
  {year} {2003})},\ \Eprint {http://arxiv.org/abs/cond-mat/0205033}
  {cond-mat/0205033} \BibitemShut {NoStop}%
\bibitem [{\citenamefont {Frahm}\ and\ \citenamefont {Martins}(2015)}]{FrMa15}%
  \BibitemOpen
  \bibfield  {author} {\bibinfo {author} {\bibfnamefont {H.}~\bibnamefont
  {Frahm}}\ and\ \bibinfo {author} {\bibfnamefont {M.~J.}\ \bibnamefont
  {Martins}},\ }\href@noop {} {\bibfield  {journal} {\bibinfo  {journal} {Nucl.
  Phys. B}\ }\textbf {\bibinfo {volume} {894}},\ \bibinfo {pages} {665}
  (\bibinfo {year} {2015})},\ \Eprint {http://arxiv.org/abs/1502.05305}
  {arXiv:1502.05305} \BibitemShut {NoStop}%
\bibitem [{\citenamefont {Frahm}\ and\ \citenamefont {Martins}(2022)}]{FrMa22}%
  \BibitemOpen
  \bibfield  {author} {\bibinfo {author} {\bibfnamefont {H.}~\bibnamefont
  {Frahm}}\ and\ \bibinfo {author} {\bibfnamefont {M.~J.}\ \bibnamefont
  {Martins}},\ }\href {\doibase 10.1016/j.nuclphysb.2022.115799} {\bibfield
  {journal} {\bibinfo  {journal} {Nucl. Phys. B}\ }\textbf {\bibinfo {volume}
  {980}},\ \bibinfo {pages} {115799} (\bibinfo {year} {2022})},\ \Eprint
  {http://arxiv.org/abs/2202.13405} {arXiv:2202.13405} \BibitemShut {NoStop}%
\bibitem [{\citenamefont {Bl{\"o}te}\ \emph {et~al.}(1986)\citenamefont
  {Bl{\"o}te}, \citenamefont {Cardy},\ and\ \citenamefont
  {Nightingale}}]{BlCN86}%
  \BibitemOpen
  \bibfield  {author} {\bibinfo {author} {\bibfnamefont {H.~W.~J.}\
  \bibnamefont {Bl{\"o}te}}, \bibinfo {author} {\bibfnamefont {J.~L.}\
  \bibnamefont {Cardy}}, \ and\ \bibinfo {author} {\bibfnamefont {M.~P.}\
  \bibnamefont {Nightingale}},\ }\href@noop {} {\bibfield  {journal} {\bibinfo
  {journal} {Phys. Rev. Lett.}\ }\textbf {\bibinfo {volume} {56}},\ \bibinfo
  {pages} {742} (\bibinfo {year} {1986})}\BibitemShut {NoStop}%
\bibitem [{\citenamefont {Alcaraz}\ \emph {et~al.}(1987)\citenamefont
  {Alcaraz}, \citenamefont {Barber}, \citenamefont {Batchelor}, \citenamefont
  {Baxter},\ and\ \citenamefont {Quispel}}]{ABBBQ87}%
  \BibitemOpen
  \bibfield  {author} {\bibinfo {author} {\bibfnamefont {F.}~\bibnamefont
  {Alcaraz}}, \bibinfo {author} {\bibfnamefont {M.}~\bibnamefont {Barber}},
  \bibinfo {author} {\bibfnamefont {M.}~\bibnamefont {Batchelor}}, \bibinfo
  {author} {\bibfnamefont {R.}~\bibnamefont {Baxter}}, \ and\ \bibinfo {author}
  {\bibfnamefont {G.}~\bibnamefont {Quispel}},\ }\href@noop {} {\bibfield
  {journal} {\bibinfo  {journal} {J. Phys. A: Math. Gen.}\ }\textbf {\bibinfo
  {volume} {20}},\ \bibinfo {pages} {6397} (\bibinfo {year}
  {1987})}\BibitemShut {NoStop}%
\bibitem [{\citenamefont {Martins}(1990)}]{Mart90}%
  \BibitemOpen
  \bibfield  {author} {\bibinfo {author} {\bibfnamefont {M.~J.}\ \bibnamefont
  {Martins}},\ }\href {\doibase 10.1016/0375-9601(90)90472-Z} {\bibfield
  {journal} {\bibinfo  {journal} {Phys. Lett. A}\ }\textbf {\bibinfo {volume}
  {151}},\ \bibinfo {pages} {519} (\bibinfo {year} {1990})}\BibitemShut
  {NoStop}%
\bibitem [{\citenamefont {{Van der Jeugt}}(1984)}]{Jeugt84}%
  \BibitemOpen
  \bibfield  {author} {\bibinfo {author} {\bibfnamefont {J.}~\bibnamefont {{Van
  der Jeugt}}},\ }\href@noop {} {\bibfield  {journal} {\bibinfo  {journal} {J.
  Math. Phys.}\ }\textbf {\bibinfo {volume} {25}},\ \bibinfo {pages} {3334}
  (\bibinfo {year} {1984})}\BibitemShut {NoStop}%
\bibitem [{\citenamefont {Cherednik}(1984)}]{Cher84}%
  \BibitemOpen
  \bibfield  {author} {\bibinfo {author} {\bibfnamefont {I.~V.}\ \bibnamefont
  {Cherednik}},\ }\href@noop {} {\bibfield  {journal} {\bibinfo  {journal}
  {Theor. Math. Phys.}\ }\textbf {\bibinfo {volume} {61}},\ \bibinfo {pages}
  {977} (\bibinfo {year} {1984})}\BibitemShut {NoStop}%
\bibitem [{\citenamefont {Sklyanin}(1988)}]{Skly88}%
  \BibitemOpen
  \bibfield  {author} {\bibinfo {author} {\bibfnamefont {E.~K.}\ \bibnamefont
  {Sklyanin}},\ }\href@noop {} {\bibfield  {journal} {\bibinfo  {journal} {J.
  Phys. A: Math. Gen.}\ }\textbf {\bibinfo {volume} {21}},\ \bibinfo {pages}
  {2375} (\bibinfo {year} {1988})}\BibitemShut {NoStop}%
\bibitem [{\citenamefont {Mezincescu}\ and\ \citenamefont
  {Nepomechie}(1991)}]{MeNe91a}%
  \BibitemOpen
  \bibfield  {author} {\bibinfo {author} {\bibfnamefont {L.}~\bibnamefont
  {Mezincescu}}\ and\ \bibinfo {author} {\bibfnamefont {R.~I.}\ \bibnamefont
  {Nepomechie}},\ }\href {\doibase 10.1142/S0217751X91002458} {\bibfield
  {journal} {\bibinfo  {journal} {Int. J. Mod. Phys. A}\ }\textbf {\bibinfo
  {volume} {6}},\ \bibinfo {pages} {5231} (\bibinfo {year} {1991})},\ \bibinfo
  {note} {addendum: Int. J. Mod. Phys. A7 (1992) 5657-5659}\BibitemShut
  {NoStop}%
\bibitem [{\citenamefont {Kulish}\ and\ \citenamefont
  {Sklyanin}(1991)}]{KuSk91}%
  \BibitemOpen
  \bibfield  {author} {\bibinfo {author} {\bibfnamefont {P.~P.}\ \bibnamefont
  {Kulish}}\ and\ \bibinfo {author} {\bibfnamefont {E.~K.}\ \bibnamefont
  {Sklyanin}},\ }\href {\doibase 10.1088/0305-4470/24/8/009} {\bibfield
  {journal} {\bibinfo  {journal} {J. Phys. A: Math. Gen.}\ }\textbf {\bibinfo
  {volume} {24}},\ \bibinfo {pages} {L435} (\bibinfo {year}
  {1991})}\BibitemShut {NoStop}%
\bibitem [{\citenamefont {Links}\ and\ \citenamefont {Gould}(1996)}]{LiGo96}%
  \BibitemOpen
  \bibfield  {author} {\bibinfo {author} {\bibfnamefont {J.~R.}\ \bibnamefont
  {Links}}\ and\ \bibinfo {author} {\bibfnamefont {M.~D.}\ \bibnamefont
  {Gould}},\ }\href {\doibase 10.1142/S0217979296001859} {\bibfield  {journal}
  {\bibinfo  {journal} {Int. J. Mod. Phys. B}\ }\textbf {\bibinfo {volume}
  {10}},\ \bibinfo {pages} {3461} (\bibinfo {year} {1996})}\BibitemShut
  {NoStop}%
\bibitem [{\citenamefont {Bracken}\ \emph {et~al.}(1998)\citenamefont
  {Bracken}, \citenamefont {Ge}, \citenamefont {Zhang},\ and\ \citenamefont
  {Zhou}}]{BGZZ98}%
  \BibitemOpen
  \bibfield  {author} {\bibinfo {author} {\bibfnamefont {A.~J.}\ \bibnamefont
  {Bracken}}, \bibinfo {author} {\bibfnamefont {X.-Y.}\ \bibnamefont {Ge}},
  \bibinfo {author} {\bibfnamefont {Y.-Z.}\ \bibnamefont {Zhang}}, \ and\
  \bibinfo {author} {\bibfnamefont {H.-Q.}\ \bibnamefont {Zhou}},\ }\href@noop
  {} {\bibfield  {journal} {\bibinfo  {journal} {Nucl. Phys. B}\ }\textbf
  {\bibinfo {volume} {516}},\ \bibinfo {pages} {588} (\bibinfo {year}
  {1998})},\ \Eprint {http://arxiv.org/abs/cond-mat/9710141} {cond-mat/9710141}
  \BibitemShut {NoStop}%
\bibitem [{\citenamefont {Galleas}\ and\ \citenamefont
  {Martins}(2004)}]{GaMa04}%
  \BibitemOpen
  \bibfield  {author} {\bibinfo {author} {\bibfnamefont {W.}~\bibnamefont
  {Galleas}}\ and\ \bibinfo {author} {\bibfnamefont {M.~J.}\ \bibnamefont
  {Martins}},\ }\href {\doibase 10.1016/j.nuclphysb.2004.08.002} {\bibfield
  {journal} {\bibinfo  {journal} {Nucl. Phys. B}\ }\textbf {\bibinfo {volume}
  {699}},\ \bibinfo {pages} {455} (\bibinfo {year} {2004})},\ \Eprint
  {http://arxiv.org/abs/nlin/0406003} {nlin/0406003} \BibitemShut {NoStop}%
\bibitem [{\citenamefont {Jones}(1990)}]{Jones90}%
  \BibitemOpen
  \bibfield  {author} {\bibinfo {author} {\bibfnamefont {V.~F.~R.}\
  \bibnamefont {Jones}},\ }\href {\doibase 10.1142/S021797929000036X}
  {\bibfield  {journal} {\bibinfo  {journal} {Int. J. Mod. Phys. B}\ }\textbf
  {\bibinfo {volume} {4}},\ \bibinfo {pages} {701} (\bibinfo {year}
  {1990})}\BibitemShut {NoStop}%
\bibitem [{\citenamefont {Mezincescu}\ and\ \citenamefont
  {Nepomechie}(1992)}]{MeNe92a}%
  \BibitemOpen
  \bibfield  {author} {\bibinfo {author} {\bibfnamefont {L.}~\bibnamefont
  {Mezincescu}}\ and\ \bibinfo {author} {\bibfnamefont {R.~I.}\ \bibnamefont
  {Nepomechie}},\ }\href {\doibase 10.1016/0550-3213(92)90367-K} {\bibfield
  {journal} {\bibinfo  {journal} {Nucl. Phys. B}\ }\textbf {\bibinfo {volume}
  {372}},\ \bibinfo {pages} {597} (\bibinfo {year} {1992})}\BibitemShut
  {NoStop}%
\bibitem [{\citenamefont {Artz}\ \emph
  {et~al.}(1995{\natexlab{a}})\citenamefont {Artz}, \citenamefont
  {Mezincescu},\ and\ \citenamefont {Nepomechie}}]{ArMN95a}%
  \BibitemOpen
  \bibfield  {author} {\bibinfo {author} {\bibfnamefont {S.}~\bibnamefont
  {Artz}}, \bibinfo {author} {\bibfnamefont {L.}~\bibnamefont {Mezincescu}}, \
  and\ \bibinfo {author} {\bibfnamefont {R.~I.}\ \bibnamefont {Nepomechie}},\
  }\href {\doibase 10.1088/0305-4470/28/18/006} {\bibfield  {journal} {\bibinfo
   {journal} {J. Phys. A: Math. Gen.}\ }\textbf {\bibinfo {volume} {28}},\
  \bibinfo {pages} {5131} (\bibinfo {year} {1995}{\natexlab{a}})},\ \Eprint
  {http://arxiv.org/abs/hep-th/9504085v2} {hep-th/9504085v2} \BibitemShut
  {NoStop}%
\bibitem [{\citenamefont {Artz}\ \emph
  {et~al.}(1995{\natexlab{b}})\citenamefont {Artz}, \citenamefont
  {Mezincescu},\ and\ \citenamefont {Nepomechie}}]{ArMN95}%
  \BibitemOpen
  \bibfield  {author} {\bibinfo {author} {\bibfnamefont {S.}~\bibnamefont
  {Artz}}, \bibinfo {author} {\bibfnamefont {L.}~\bibnamefont {Mezincescu}}, \
  and\ \bibinfo {author} {\bibfnamefont {R.~I.}\ \bibnamefont {Nepomechie}},\
  }\href {\doibase 10.1142/S0217751X95000942} {\bibfield  {journal} {\bibinfo
  {journal} {Int. J. Mod. Phys. A}\ }\textbf {\bibinfo {volume} {10}},\
  \bibinfo {pages} {1937} (\bibinfo {year} {1995}{\natexlab{b}})},\ \Eprint
  {http://arxiv.org/abs/hep-th/9409130} {hep-th/9409130} \BibitemShut {NoStop}%
\bibitem [{\citenamefont {Yung}\ and\ \citenamefont
  {Batchelor}(1995)}]{YuBa95c}%
  \BibitemOpen
  \bibfield  {author} {\bibinfo {author} {\bibfnamefont {C.~M.}\ \bibnamefont
  {Yung}}\ and\ \bibinfo {author} {\bibfnamefont {M.~T.}\ \bibnamefont
  {Batchelor}},\ }\href {\doibase 10.1016/0375-9601(95)00083-F} {\bibfield
  {journal} {\bibinfo  {journal} {Phys. Lett. A}\ }\textbf {\bibinfo {volume}
  {198}},\ \bibinfo {pages} {395} (\bibinfo {year} {1995})},\ \Eprint
  {http://arxiv.org/abs/hep-th/9502039} {hep-th/9502039} \BibitemShut {NoStop}%
\bibitem [{\citenamefont {Essler}\ and\ \citenamefont
  {Korepin}(1992)}]{EsKo92}%
  \BibitemOpen
  \bibfield  {author} {\bibinfo {author} {\bibfnamefont {F.~H.~L.}\
  \bibnamefont {Essler}}\ and\ \bibinfo {author} {\bibfnamefont {V.~E.}\
  \bibnamefont {Korepin}},\ }\href@noop {} {\bibfield  {journal} {\bibinfo
  {journal} {Phys. Rev. B}\ }\textbf {\bibinfo {volume} {46}},\ \bibinfo
  {pages} {9147} (\bibinfo {year} {1992})}\BibitemShut {NoStop}%
\bibitem [{\citenamefont {Yang}\ and\ \citenamefont {Yang}(1969)}]{YaYa69}%
  \BibitemOpen
  \bibfield  {author} {\bibinfo {author} {\bibfnamefont {C.~N.}\ \bibnamefont
  {Yang}}\ and\ \bibinfo {author} {\bibfnamefont {C.~P.}\ \bibnamefont
  {Yang}},\ }\href@noop {} {\bibfield  {journal} {\bibinfo  {journal} {J. Math.
  Phys.}\ }\textbf {\bibinfo {volume} {10}},\ \bibinfo {pages} {1115} (\bibinfo
  {year} {1969})}\BibitemShut {NoStop}%
\bibitem [{\citenamefont {Pasquier}\ and\ \citenamefont
  {Saleur}(1990)}]{PaSa90}%
  \BibitemOpen
  \bibfield  {author} {\bibinfo {author} {\bibfnamefont {V.}~\bibnamefont
  {Pasquier}}\ and\ \bibinfo {author} {\bibfnamefont {H.}~\bibnamefont
  {Saleur}},\ }\href@noop {} {\bibfield  {journal} {\bibinfo  {journal} {Nucl.
  Phys. B}\ }\textbf {\bibinfo {volume} {330}},\ \bibinfo {pages} {523}
  (\bibinfo {year} {1990})}\BibitemShut {NoStop}%
\bibitem [{\citenamefont {Batchelor}\ \emph {et~al.}(1990)\citenamefont
  {Batchelor}, \citenamefont {Mezincescu}, \citenamefont {Nepomechie},\ and\
  \citenamefont {Rittenberg}}]{BMNR90}%
  \BibitemOpen
  \bibfield  {author} {\bibinfo {author} {\bibfnamefont {M.~T.}\ \bibnamefont
  {Batchelor}}, \bibinfo {author} {\bibfnamefont {L.}~\bibnamefont
  {Mezincescu}}, \bibinfo {author} {\bibfnamefont {R.~I.}\ \bibnamefont
  {Nepomechie}}, \ and\ \bibinfo {author} {\bibfnamefont {V.}~\bibnamefont
  {Rittenberg}},\ }\href {\doibase 10.1088/0305-4470/23/4/003} {\bibfield
  {journal} {\bibinfo  {journal} {J. Phys. A: Math. Gen.}\ }\textbf {\bibinfo
  {volume} {23}},\ \bibinfo {pages} {L141} (\bibinfo {year}
  {1990})}\BibitemShut {NoStop}%
\bibitem [{\citenamefont {Zamolodchikov}\ and\ \citenamefont
  {Fateev}(1980)}]{ZaFa80}%
  \BibitemOpen
  \bibfield  {author} {\bibinfo {author} {\bibfnamefont {A.~B.}\ \bibnamefont
  {Zamolodchikov}}\ and\ \bibinfo {author} {\bibfnamefont {V.~A.}\ \bibnamefont
  {Fateev}},\ }\href@noop {} {\bibfield  {journal} {\bibinfo  {journal} {Sov.
  J. Nucl. Phys.}\ }\textbf {\bibinfo {volume} {32}},\ \bibinfo {pages} {298}
  (\bibinfo {year} {1980})}\BibitemShut {NoStop}%
\bibitem [{\citenamefont {Mezincescu}\ \emph {et~al.}(1990)\citenamefont
  {Mezincescu}, \citenamefont {Nepomechie},\ and\ \citenamefont
  {Rittenberg}}]{MeNR90}%
  \BibitemOpen
  \bibfield  {author} {\bibinfo {author} {\bibfnamefont {L.}~\bibnamefont
  {Mezincescu}}, \bibinfo {author} {\bibfnamefont {R.~I.}\ \bibnamefont
  {Nepomechie}}, \ and\ \bibinfo {author} {\bibfnamefont {V.}~\bibnamefont
  {Rittenberg}},\ }\href {\doibase 10.1016/0375-9601(90)90016-H} {\bibfield
  {journal} {\bibinfo  {journal} {Phys. Lett. A}\ }\textbf {\bibinfo {volume}
  {147}},\ \bibinfo {pages} {70} (\bibinfo {year} {1990})}\BibitemShut
  {NoStop}%
\bibitem [{\citenamefont {di~Francesco}\ \emph {et~al.}(1988)\citenamefont
  {di~Francesco}, \citenamefont {Saleur},\ and\ \citenamefont
  {Zuber}}]{FrSZ88}%
  \BibitemOpen
  \bibfield  {author} {\bibinfo {author} {\bibfnamefont {P.}~\bibnamefont
  {di~Francesco}}, \bibinfo {author} {\bibfnamefont {H.}~\bibnamefont
  {Saleur}}, \ and\ \bibinfo {author} {\bibfnamefont {J.-B.}\ \bibnamefont
  {Zuber}},\ }\href {\doibase 10.1016/0550-3213(88)90605-0} {\bibfield
  {journal} {\bibinfo  {journal} {Nucl. Phys. B}\ }\textbf {\bibinfo {volume}
  {300}},\ \bibinfo {pages} {393} (\bibinfo {year} {1988})}\BibitemShut
  {NoStop}%
\bibitem [{\citenamefont {Alcaraz}\ and\ \citenamefont
  {Martins}(1989)}]{AlMa89}%
  \BibitemOpen
  \bibfield  {author} {\bibinfo {author} {\bibfnamefont {F.~C.}\ \bibnamefont
  {Alcaraz}}\ and\ \bibinfo {author} {\bibfnamefont {M.~J.}\ \bibnamefont
  {Martins}},\ }\href {\doibase 10.1088/0305-4470/22/11/023} {\bibfield
  {journal} {\bibinfo  {journal} {J. Phys. A: Math. Gen.}\ }\textbf {\bibinfo
  {volume} {22}},\ \bibinfo {pages} {1829} (\bibinfo {year}
  {1989})}\BibitemShut {NoStop}%
\bibitem [{\citenamefont {Alcaraz}\ and\ \citenamefont
  {Martins}(1990)}]{AlMa90}%
  \BibitemOpen
  \bibfield  {author} {\bibinfo {author} {\bibfnamefont {F.~C.}\ \bibnamefont
  {Alcaraz}}\ and\ \bibinfo {author} {\bibfnamefont {M.~J.}\ \bibnamefont
  {Martins}},\ }\href@noop {} {\bibfield  {journal} {\bibinfo  {journal} {J.
  Phys. A: Math. Gen.}\ }\textbf {\bibinfo {volume} {23}},\ \bibinfo {pages}
  {1439} (\bibinfo {year} {1990})}\BibitemShut {NoStop}%
\bibitem [{\citenamefont {Frahm}\ \emph {et~al.}(1990)\citenamefont {Frahm},
  \citenamefont {Yu},\ and\ \citenamefont {Fowler}}]{FrYF90}%
  \BibitemOpen
  \bibfield  {author} {\bibinfo {author} {\bibfnamefont {H.}~\bibnamefont
  {Frahm}}, \bibinfo {author} {\bibfnamefont {N.-C.}\ \bibnamefont {Yu}}, \
  and\ \bibinfo {author} {\bibfnamefont {M.}~\bibnamefont {Fowler}},\
  }\href@noop {} {\bibfield  {journal} {\bibinfo  {journal} {Nucl. Phys. B}\
  }\textbf {\bibinfo {volume} {336}},\ \bibinfo {pages} {396} (\bibinfo {year}
  {1990})}\BibitemShut {NoStop}%
\bibitem [{\citenamefont {Koo}\ and\ \citenamefont {Saleur}(1993)}]{KoSa93}%
  \BibitemOpen
  \bibfield  {author} {\bibinfo {author} {\bibfnamefont {W.~M.}\ \bibnamefont
  {Koo}}\ and\ \bibinfo {author} {\bibfnamefont {H.}~\bibnamefont {Saleur}},\
  }\href {\doibase 10.1142/S0217751X93002071} {\bibfield  {journal} {\bibinfo
  {journal} {Int. J. Mod. Phys. A}\ }\textbf {\bibinfo {volume} {8}},\ \bibinfo
  {pages} {5165} (\bibinfo {year} {1993})},\ \Eprint
  {http://arxiv.org/abs/hep-th/9303118} {hep-th/9303118} \BibitemShut {NoStop}%
\bibitem [{\citenamefont {von Gehlen}\ and\ \citenamefont
  {Rittenberg}(1986)}]{GeRi86a}%
  \BibitemOpen
  \bibfield  {author} {\bibinfo {author} {\bibfnamefont {G.}~\bibnamefont {von
  Gehlen}}\ and\ \bibinfo {author} {\bibfnamefont {V.}~\bibnamefont
  {Rittenberg}},\ }\href {\doibase 10.1088/0305-4470/19/10/014} {\bibfield
  {journal} {\bibinfo  {journal} {J. Phys. A: Math. Gen.}\ }\textbf {\bibinfo
  {volume} {19}},\ \bibinfo {pages} {L631} (\bibinfo {year}
  {1986})}\BibitemShut {NoStop}%
\bibitem [{\citenamefont {Cardy}(1986)}]{Card86a}%
  \BibitemOpen
  \bibfield  {author} {\bibinfo {author} {\bibfnamefont {J.~L.}\ \bibnamefont
  {Cardy}},\ }\href {\doibase 10.1016/0550-3213(86)90552-3} {\bibfield
  {journal} {\bibinfo  {journal} {Nucl. Phys. B}\ }\textbf {\bibinfo {volume}
  {270}},\ \bibinfo {pages} {186} (\bibinfo {year} {1986})}\BibitemShut
  {NoStop}%
\bibitem [{\citenamefont {Lima-Santos}(2009)}]{Lima09}%
  \BibitemOpen
  \bibfield  {author} {\bibinfo {author} {\bibfnamefont {A.}~\bibnamefont
  {Lima-Santos}},\ }\href {\doibase 10.1088/1742-5468/2009/07/P07045}
  {\bibfield  {journal} {\bibinfo  {journal} {J. Stat. Mech.}\ ,\ \bibinfo
  {pages} {P07045}} (\bibinfo {year} {2009})},\ \Eprint
  {http://arxiv.org/abs/0809.0421} {arXiv:0809.0421} \BibitemShut {NoStop}%
\bibitem [{\citenamefont {Frahm}\ and\ \citenamefont
  {Martins}(2023)}]{FrMa23.data}%
  \BibitemOpen
  \bibfield  {author} {\bibinfo {author} {\bibfnamefont {H.}~\bibnamefont
  {Frahm}}\ and\ \bibinfo {author} {\bibfnamefont {M.~J.}\ \bibnamefont
  {Martins}},\ }\href {\doibase 10.25835/ypipefbz} {\enquote {\bibinfo {title}
  {{Dataset: Finite size data for $U_q[OSp(3|2)]$ quantum chains with quantum
  group invariant boundary conditions}},}\ }\bibinfo {howpublished}
  {{\url{https://doi.org/10.25835/ypipefbz}}} (\bibinfo {year} {2023}),\
  \bibinfo {note} {{Research Data Repository, Leibniz Universit\"at
  Hannover}}\BibitemShut {NoStop}%
\end{thebibliography}

%

\end{document}